\documentclass[12pt, letterpaper]{article}
\usepackage{siunitx} 
\usepackage[utf8]{inputenc}
\usepackage{amsmath, amssymb, amsfonts}  %
\usepackage{graphicx}
\usepackage[margin=1in]{geometry}
\usepackage{natbib}  
\usepackage{appendix}
\usepackage[raggedright]{titlesec}
\usepackage[hang,flushmargin]{footmisc} 
\usepackage{color}
\usepackage[svgnames]{xcolor}  
\usepackage{yfonts}  
\usepackage{pifont}  
\usepackage{multirow}  
\usepackage{subfigure}  
\usepackage{caption}  
\usepackage{threeparttable}  
\usepackage{esvect}  
\usepackage{tikz, tikz-cd}  
\usetikzlibrary{positioning, fadings, patterns, shadows.blur, shapes}  
\usepackage{bbm}  
\usepackage{colortbl}  
\usepackage{pdflscape}  
\usepackage{booktabs}  
\usepackage{amsmath}  
\usepackage{mathdots}  
\usepackage{yhmath}  
\usepackage{cancel}  
\usepackage{array}  
\usepackage{tabularx}  
\usepackage{makecell}
\usepackage{placeins}  
\usepackage{float}
\usepackage{comment}  
\usepackage[ruled,linesnumbered,commentsnumbered]{algorithm2e}  
\usepackage[font=small,labelfont=bf]{caption}  

\usepackage{hyperref}  
\hypersetup{
    colorlinks=true,
    linkcolor=MidnightBlue,
    filecolor=magenta,      
    urlcolor=cyan,
    citecolor=MidnightBlue
}

\usepackage{rotating}  
\usepackage{ntheorem}  
\usepackage{etoc}   
\makeatletter
\renewtheoremstyle{plain}{\item{\theorem@headerfont ##1\ ##2\theorem@separator}~}{\item{\theorem@headerfont ##1\ ##2\ (##3)\theorem@separator}~}
\makeatother

\newcommand{\cor}{\text{cor}}

\newcommand{\bx}{\mathbf{x}}

\newcommand{\by}{\mathbf{y}}
\newcommand{\bZ}{\mathbf{Z}}
\newcommand{\bZh}{\hat{\mathbf{Z}}}
\newcommand{\bz}{\mathbf{z}}
\newcommand{\bzh}{\hat{\mathbf{z}}}
\newcommand{\bD}{\mathbf{D}}

\newcommand{\bW}{\mathbf{W}}
\newcommand{\bWh}{\hat{\mathbf{W}}}

\newcommand{\R}{\mathbbm{R}}

\definecolor{shadecolor}{gray}{0.9}

\DeclareMathOperator*{\argmin}{arg\,min}

\tikzset{every picture/.style={line width=0.75pt}} 

\usepackage{enumitem}  
\newlist{Step}{enumerate}{2}
\setlist[Step]{label={{Step \arabic*.}}, leftmargin=*}
\usepackage{setspace}  

\makeatletter
\newcommand\circled[1]{%
  \mathpalette\@circled{#1}%
}
\newcommand\@circled[2]{%
  \tikz[baseline=(math.base)] \node[draw,circle,inner sep=2pt] (math) {$\m@th#1#2$};%
}
\newcommand\circledblue[1]{%
  \mathpalette\@circledblue{#1}%
}
\newcommand\@circledblue[2]{%
  \tikz[baseline=(math.base)] \node[draw,circle, fill=blue!20, inner sep=2pt] (math) {$\m@th#1#2$};%
}
\makeatother

\renewenvironment{abstract}
 {\begin{center}\normalsize\textsc{Abstract}%
 \end{center}\begin{quote}\normalsize}
 {\end{quote}}

\usepackage{apptools}  
\AtAppendix{\counterwithin{assumption}{section}}

\makeatletter
\newcommand{\myfnsymbol}[1]{%
  \expandafter\@myfnsymbol\csname c@#1\endcsname
}
\newcommand{\@myfnsymbol}[1]{%
  \ifcase #1
  \or $^\dagger$
  \or $^*$
  \or 1
  \or 2
  \or 3
  \fi
}
\newcommand{\acknowledgements}{\@myfnsymbol{1}}
\newcommand{\equalcontributor}{\@myfnsymbol{2}}
\newcommand{\affiliationA}{\@myfnsymbol{3}}
\newcommand{\affiliationB}{\@myfnsymbol{4}}
\newcommand{\affiliationC}{\@myfnsymbol{5}}
\makeatother

\bibpunct[: ]{(}{)}{;}{a}{}{,~} 

\title{Estimating Consensus Ideal Points Using Multi-Source Data\thanks{The authors thank Scott Abramson, Josh Clinton, Josh Kalla, and participants at PolMeth 2025 for their feedback and comments.}}
\author{Mellissa Meisels\footnote{Assistant Professor, Department of Political Science, Yale University, Email: \texttt{mellissa.meisels@yale.edu}, Website: \texttt{http://www.mellissameisels.com}} \and Melody Huang\footnote{Assistant Professor, Department of Political Science and Statistics \& Data Science, Yale University, Email: \texttt{melody.huang@yale.edu}, Website: \texttt{http://www.melodyyhuang.com}} \and Tiffany M. Tang\footnote{Clare Boothe Luce Assistant Professor, Department of Applied and Computational Mathematics and Statistics, University of Notre Dame, Email: \texttt{ttang4@nd.edu}, Website: \texttt{http://tiffanymtang.github.io}}}
\date{}

\begin{document}

\maketitle

\begin{abstract} 
In the advent of big data and machine learning, researchers now have a wealth of congressional candidate ideal point estimates at their disposal for theory testing. Weak relationships raise questions about the extent to which they capture a shared quantity --- rather than idiosyncratic, domain--specific factors --- yet different measures are used interchangeably in most substantive analyses. Moreover, questions central to the study of American politics implicate relationships between candidate ideal points and other variables derived from the same data sources, introducing endogeneity. We propose a method, consensus multidimensional scaling (CoMDS), which better aligns with how applied scholars use ideal points in practice. CoMDS captures the shared, stable associations of a set of underlying ideal point estimates and can be interpreted as their common spatial representation. We illustrate the utility of our approach for assessing relationships within domains of existing measures and provide a suite of diagnostic tools to aid in practical usage.

\end{abstract} 

\clearpage 
\doublespacing
\section{Introduction} 
Measures of candidates' positions are the fundamental building blocks for testing theories related to congressional polarization, accountability, and representation \citep{canes-wrone_out_2002, clinton_representation_2006, hall_what_2015, mccarty_polarized_2006}. Perhaps the most ubiquitous and longstanding is NOMINATE, a roll--call based estimate of sitting legislators' ideal points \citep{poole2011ScalingRollCall}. The advent of big data and machine learning has introduced new  approaches to constructing ideal points in political science. Examples of recently proposed approaches include transaction--level campaign finance receipts (e.g., \citealp{bonica2014mapping, bonica2018rollcall, bonica_database_2024}), campaign website platforms \citep{meisels_candidate_2025}, Tweets \citep[e.g.,][]{barbera2015birds, cowburn2025partisan}, Facebook posts \citep[e.g.,][]{bond2015quantifying}, legislative speeches \citep[e.g.,][]{lauderdale2016measuring}, and most recently, LLM-generated embeddings \citep{burnham2024semantic}.

Different ideal point estimates depend on not only the specific data sources researchers have leveraged, but also the behavioral assumptions and estimation approaches used to construct latent factor models from the underlying data. The extent to which various ideal point estimates relate to one another is an open question both substantively and methodologically: recent evidence suggests that measures are very weakly related within party \citep{barber_comparing_2022, meisels_candidate_2025, tausanovitch_estimating_2017}.  While all seek to capture the common concept of a candidate's ideal point along a liberal--conservative spectrum, in reality, each is likely measured with a considerable amount of domain--specificity as well. As a result, an estimated ideal point will be an unidentifiable mixture of both a \textit{true} ideal point\footnote{We refer to a maximally general and agnostic conception of a \textit{true} ideal point as one that simply presents itself irrespective of particular institutional contexts. We remain agnostic about whether such ideal points represent candidates' deeply--held ideologies, strategic positions, or even partisan strength.} and an idiosyncratic component that is domain and context specific. 

In practice, applied researchers are typically interested in a more general latent concept of ``ideology" or ``positioning", and are agnostic about specific estimates (and the data and approach used therein). Instead, questions such as whether candidates are still responsive to their districts, how extremism impacts candidates' electoral and fundraising performances, and how different nominating institutions affect the composition of the candidate field implicate relatively general conceptualizations of candidate ``ideology." Despite the aforementioned studies raising questions about whether different measures capture a common latent concept and related calls to tailor measures to specific applications (e.g., \citealt{caughey_substance_2016}), scholars engaging in substantive theory--testing continue to use such ideal point estimates more-or-less interchangeably, ignoring contextual differences.

In this paper, we propose a consensus-based approach which provides a closer approximation of how ideal point estimates tend to be used in practice. The method, \textit{consensus multidimensional scaling} (CoMDS), aggregates a set of ideal point estimates across different data sources and estimation approaches to construct a \textit{consensus} ideal point estimate. This captures the shared, stable associations across source ideal point estimates, and can be interpreted as their common spacial representation. To help researchers identify the extent to which different substantive conclusions about U.S. congressional politics and elections may be specific to idiosyncratic features of specific data sources, we further introduce a projection-based approach to decompose each source ideal point estimate into two orthogonal components: (1) the shared component that is captured by the consensus ideal point estimate, and (2) the remaining idiosyncratic component, not captured by the consensus ideal point estimate.

Unlike existing methods in estimating ideal points with different data sources \citep[e.g.,][]{enamoradoScalingDataMultiple2021}, our proposed method is agnostic to the underlying behavioral model and data-generating assumptions for the source ideal point estimates, allowing researchers to encode in domain-specific substantive priors for each source ideal point estimate. Furthermore, unlike alternative aggregation methods like principal component analysis (PCA), CoMDS is robust to rotations, rescalings, and shifts in the source ideal point estimates, and can accommodate settings in which different data sources have differential amounts of missingness, allowing for a large degree of flexibility in methodological implementation. 

We elucidate the utility of CoMDS with an application to congressional candidate ideal points. In addition to the aforementioned weak relationships between existing measures, another problem is that many research questions implicate candidate ideal points  and a key variable on the other side of the equation which are both drawn from the same data source. We investigate relationships of substantive interest within the domains of source measures, demonstrating first that relying on domain--specific ideal points may lead to conclusions which are overstated at best and incorrectly signed at worst. Moreover, we show that in the absence of our consensus approach, assessing robustness of relationships across existing measures leads to results which conflict not only in magnitude but sometimes in sign. In contrast, CoMDS allows researchers to draw meaningful conclusions about how the ideal point component which is common across existing measures relates to variables of interest.

The paper proceeds as follows. In Section 2, we provide an overview of ideal point estimation in political science and review several approaches commonly used in studying congressional candidate positioning. Section 3 introduces the proposed method, consensus multidimensional scaling, and provides a suite of interpretability and stability tools to aid in practical usage. In Section 4, we present consensus estimates of candidates' ideal points, compare them to original source measures, and re-assess relationships of substantive interest within domains of existing measures. Section 5 concludes.

\section{Ideal point estimation in political science}\label{sec:ideal_points}

Ideal point estimation interprets a given set of observed data points as being generated from an underlying behavioral model, which is a function of a latent ideal point. For example, NOMINATE assumes that a legislator's probability of voting Yea versus Nay is a function of the distance between her ideal point and the two alternative policies which each voting option represents. Researchers then infer the ideal points which maximize the likelihood of an observed roll--call matrix. 

Formally, for a given source of data $Y^{(s)}$, ideal point estimates $\bZ^{(s)}\in \R^{n_s \times r_s}$ are estimated by assuming an underlying generative model: 
\begin{equation} 
    Y^{(s)} = g_s(\bZ^{(s)}) + \varepsilon^{(s)},
    \label{eqn:idealpoint_model}
\end{equation} 
where different assumptions about the underlying data generating process map to different functional forms of $g_s(\cdot)$. For example, in the context of roll-call votes, $g_s(\cdot)$ is a logistic function that maps to a legislator's utility, which dictates whether they vote Yea or Nay.

Notably, the choice of source data and estimation assumptions can result in very different ideal point estimates. For example, consider differences between roll-call-based NOMINATE versus ideal points estimated from campaign platform text \citep[e.g.,][]{meisels_candidate_2025}. Legislative roll--call votes are taken in an institutional setting which is relatively opaque to the public. In contrast, the same legislators' campaign platforms are explicitly public--facing for purposes of electioneering. With regard to agenda control, candidates are virtually unconstrained in the issues and positions which can be articulated in campaign platforms, whereas they may only vote upon the issues and in support or opposition to the proposed policies which reach the floor in Congress. Even two campaign--based data sources --- campaign platforms versus campaign contributions \citep[e.g.,][]{bonica2014mapping} --- differ critically in whether the activity is performed by the campaign itself or is instead the observed behavior of an another actor. 

Consequently, in the absence of restrictive assumptions, we generally do not expect ideal points estimated from one source of data to be equivalent to another estimated using a different source of data (i.e., $\bZ^{(s)} \neq \bZ^{(s')}$). However, different measures of candidates’ positions consistently separate Democrats from Republicans even without the inclusion of any covariates in measurement models. As such, despite relatively weak intraparty correlations, pooled correlations between different ideal point estimates remain exceptionally strong \citep{barber_comparing_2022, meisels_candidate_2025, tausanovitch_estimating_2017}. This suggests that different measures do capture some common variation. In other words, the estimated ideal points $\bZ^{(s)}$ contain both information about the underlying latent ideal point $\bZ^*$, as well as idiosyncratic aspects of the source data and functional form choices. 

To formalize, we rewrite the model in Equation \eqref{eqn:idealpoint_model} as: 
\begin{equation}
Y^{(s)} = g_s(\bZ^{*} + \mathbf{\nu}^{(s)}) + \varepsilon^{(s)},
\label{eqn:decomp}
\end{equation} 
where we have decomposed the ideal point estimates $\bZ^{(s)}$ into (1) $\bZ^*$, which represents some hypothetical, \textit{true} ideal point, and (2) $\mathbf{\nu}^{(s)}$, which represents idiosyncratic aspects specific to the domain of the source ideal point estimate.

Reconciling different methods and approaches for the estimation of ideal points can be challenging, to put it lightly. Existing work has proposed pooling  outcome data from different sources to fit a joint ideal point estimation model \citep[e.g.,][]{treier2008democracy, murray2013bayesian, quinn2004bayesian}. Unfortunately, naive pooling will result in biased estimates when the two datasets have varying amounts of observations and information \citep[e.g.,][]{jessee2016can}. This problem is especially pronounced in settings where target populations differ \citep[e.g.,][]{lewis2015does, tausanovitch2013measuring, shor2011ideological}. 

Alternative approaches have introduced data-adaptive  ways to re-weight different data sources. However, these approaches rely on strong parametric assumptions on the observed outcome data. For example, \citet{enamoradoScalingDataMultiple2021} implicitly assume the different outcomes are normally distributed, and can be linearly decomposed into a shared ideal point estimate and an idiosyncratic term. In the context of Equation \eqref{eqn:decomp}, this implies that every function $g_s(\cdot)$ for all data sources is linear in nature. In practice, researchers would not generally believe that the different behavioral models that generate the observed data would share the same functional form (nor that they are necessarily linear with respect to the ideal point estimate). Furthermore, common settings in practice use data that are binary (i.e., roll call votes) or categorical (i.e., word counts), neither of which are normally distributed. We provide further discussion about the relationship between existing methods and our proposed method in Appendix \ref{app:ted}.

In the following section, we introduce a method, \textit{consensus multidimensional scaling} (CoMDS) \citep{an2025consensus}, which estimates the shared component of a set of source ideal point estimates, which we refer to as the \textit{consensus ideal points}. Unlike existing approaches, CoMDS takes in the \textit{source-specific ideal point estimates} $\bZ^{(s)}$  to construct a consensus ideal point estimate. Because CoMDS is an unsupervised method, it does not rely on jointly modeling all of the different observed outcomes and can be used without access to the underlying outcomes. This is advantageous for several reasons. First, it allows researchers to preserve their context-specific underlying assumptions on the different behavioral models for each source of data. As a result, the outcomes for each source $Y^{(s)}$ can continue to be mapped to their own, unique behavioral model $g_s(\cdot)$. Second, it is more computationally efficient, allowing researchers to directly estimate a consensus ideal point estimate using pre-existing source ideal points. In Section 4, we apply CoMDS to congressional candidates and demonstrate how consensus ideal points allows for robust investigation of substantive relationships when variables of interest are derived from the same data as existing source measures.

\section{Consensus ideal point estimation with CoMDS} \label{sec:coMDS}

In the following section, we introduce our proposed approach, consensus multidimensional scaling (CoMDS). In Section 3.1, we formalize the optimization problem behind CoMDS and provide intuition for the estimated consensus ideal point estimate. In Section 3.2, we propose a projection-based approach for researchers to decompose the source ideal point estimates into the consensus ideal point estimate and an idiosyncratic factor. Finally, Section 3.3 provides two diagnostic tools for interpreting and using CoMDS in practice: (1) a relative error measure that estimates the individual contributions of each source ideal point estimate on the consensus ideal point estimate, and (2) a stability analysis that evaluates the sensitivity of the consensus ideal point estimate to the inclusion of the different source ideal point estimates.

\subsection{Consensus multidimensional scaling (CoMDS)}
\begin{figure} 
\centering 
\fcolorbox{black}{white}{
\includegraphics[width=0.95\textwidth, trim={0.5cm 3cm 0.5cm 0.5cm},clip]{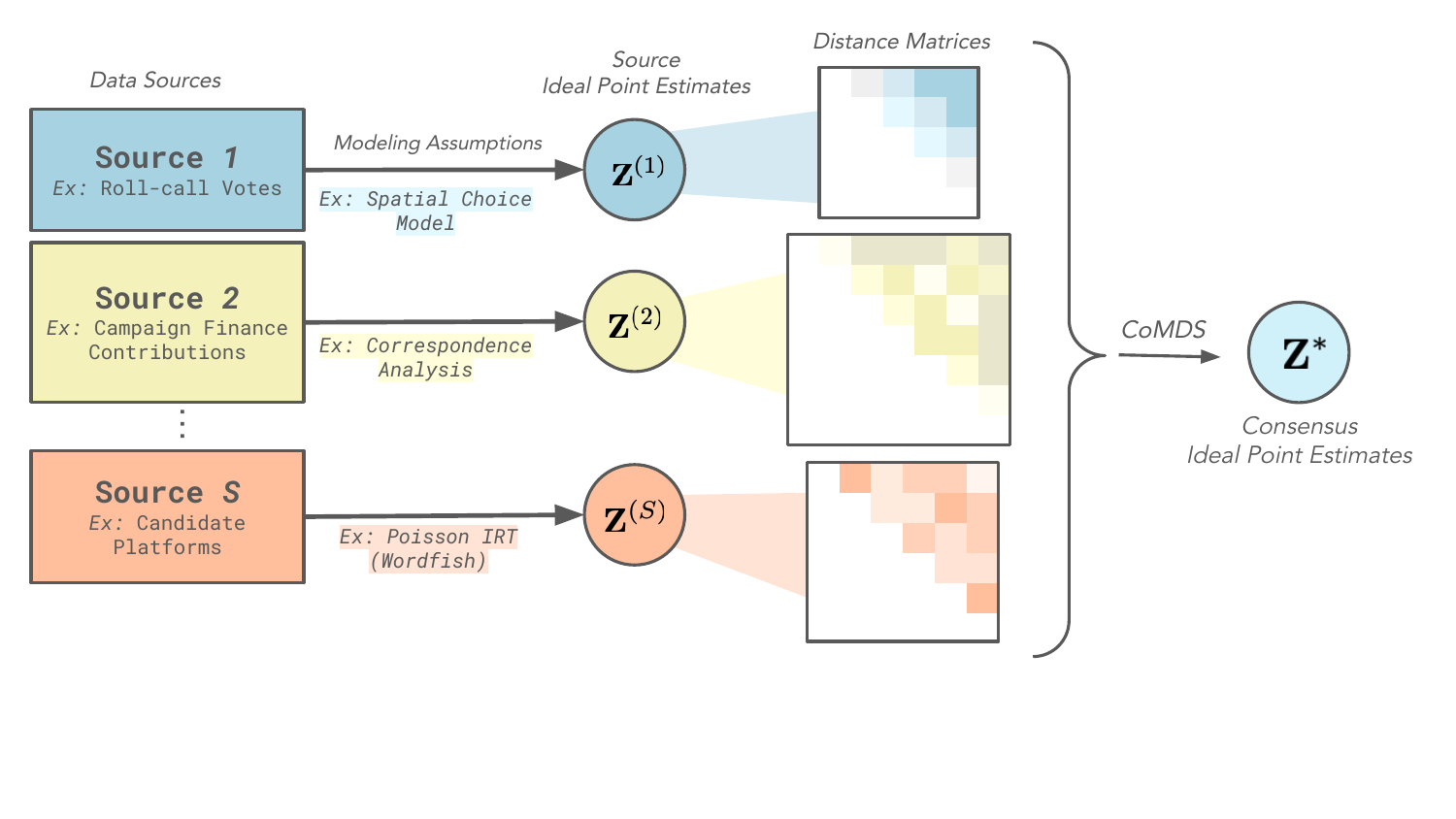}}
\caption{Overview of consensus ideal point estimation workflow}
\label{fig:schematic}
\end{figure}

At a high-level, consensus multidimensional scaling (CoMDS) takes in different source ideal point estimates $\bZ^{(s)}$ as input, evaluates the (dis)similarity between each pair of observations by computing a pairwise distance matrix for each source $\bZ^{(s)}$, and outputs a consensus ideal point estimate $\bZh^*$ that best preserves the pairwise distances observed in the source ideal point estimates \citep{an2025consensus}. Put differently, we can think of the consensus ideal point estimates from CoMDS as best approximating the shared spatial representation of underlying source ideal point estimates. Figure \ref{fig:schematic} visualizes the consensus ideal point estimation workflow. 

More formally, suppose we have $n$ candidates and $S$ source ideal point estimation approaches under study. For each $s \in \{1, ..., S\}$, let $\bZ^{(s)} \in \R^{n \times r_s}$ represent the estimated source ideal points from the $s^{th}$ ideal point estimation approach. Note that each source ideal point estimate $\bZ^{(s)}$ is allowed to have a varying number of dimensions $r_s$ and possible missing values. Without loss of generality, assume the candidates in each source have been aligned so that the $i^{th}$ candidate in source $s$ (denoted $\bz^{(s)}_i$) corresponds to the $i^{th}$ candidate in source $s'$ (denoted $\bz^{(s')}_i$). Define the missingness indicator $\alpha_{i}^{(s)}$, where $\alpha_i^{(s)} = 0$ if the $s^{th}$ source ideal point estimate is missing for candidate $i$ and $1$ otherwise. Also, let $\text{Diag}(r)$ denote the set of diagonal matrices of size $r \times r$. Using the source ideal points $\bZ^{(1)}, \ldots, \bZ^{(S)}$, CoMDS estimates the consensus ideal points $\bZ^{*}$ via the following two step procedure.

\paragraph{Step 1: Compute dissimilarity between ideal points in each source.} For each source $s = 1, \ldots, S$ and each pair of candidates $i, j = 1, \ldots, n$, evaluate the pairwise distance (or dissimilarity) between the ideal points in the $s^{th}$ source space: $D_{ij}^{(s)} := d_{ij}(\bz_i^{(s)}, \bz_j^{(s)})$, where $d(\bx, \by)$ corresponds to an arbitrary distance measure. For our purposes, we focus throughout the paper on a Euclidean distance metric (i.e., $d(\bx, \by) = \lVert \bx - \by \rVert_2$). 

\paragraph{Step 2: Estimate consensus structure across sources.} Using these pairwise distances $D_{ij}^{(s)}$, CoMDS leverages a generalization of multidimensional scaling (see Appendix~\ref{app:mds} for additional discussion), which was originally developed in the psychometrics literature \citep{carroll1970analysis}, in order to estimate the \textit{consensus ideal points} $\bZh^*$ via
\begin{align}
    \bZh^*, \bWh^{(1)}, \ldots, \bWh^{(S)} = \argmin_{\substack{\bZ \in \R^{n \times r}\\ \bW^{(s)} \in \text{Diag}(r)}} \sum_{s = 1}^{S} \sum_{i < j} \underbrace{\alpha_{i}^{(s)} \alpha_{j}^{(s)}}_{\substack{\text{missingness}\\\text{indicators}}} \Big\{ \underbrace{D_{ij}^{(s)}}_{\substack{\text{distance in}\\\text{source space}}} - \underbrace{d(\bW^{(s)} \bz_i, \bW^{(s)} \bz_j)}_{\substack{\text{distance in}\\\text{consensus space}}} \Big\}^2,\label{eqn:optim}
\end{align}
where $r$ is a pre-specified number of dimensions for the consensus output (typically, $r = 1$ or $2$), $\bWh^{(1)}, \ldots, \bWh^{(S)}$ are $r \times r$ diagonal matrices acting as source-specific weights, and $\alpha_{i}^{(s)} \alpha_{j}^{(s)}$ is a missingness indicator equaling 1 if the $s^{th}$ source ideal points for candidates $i$ and $j$ exist and 0 if either is missing.

Intuitively, CoMDS ensures that the pairwise distances between two candidates in the original source ideal point space are similar to the pairwise distances between those two candidates in the newly-learned CoMDS ideal point space. This means that the consensus ideal point from CoMDS will span the variation of the source ideal points, thereby serving as a measure of the shared associations across the different source ideal point measures. 

To solve the CoMDS optimization problem given in Equation \eqref{eqn:optim}, we use SMACOF \citep{de2009multidimensional}, which leverages an iterative majorization-minimization optimization scheme with a closed-form constrained update to minimize the objective in Equation \eqref{eqn:optim}. This optimization problem is known to converge linearly \citep{de1988convergence}.\footnote{As a reference, for a rank-1 problem with 2000 candidates and 3 sources, CoMDS took around 2 minutes to converge with a tolerance of 1E-6 using a Macbook Pro with an Apple M3 Pro chip.}

There are several advantages to CoMDS over alternative aggregation approaches. First, unlike other methods such as MD2S \citep{enamoradoScalingDataMultiple2021} and PCA \citep{bonica_database_2024}, CoMDS does not explicitly make any linearity assumptions. Second, because CoMDS operates on the pairwise distances between source ideal point estimates and allows for source-specific weight matrices $\bW^{(s)}$, it will be invariant to underlying rotations, rescalings, and shifts in the source ideal point estimates. Importantly, such transformations do not fundamentally change the interpretation of the source ideal point estimates and thus should not change the consensus ideal point estimates. However, methods such as PCA, which operate directly on the source ideal point estimates $[\bZ^{(1)}, \ldots, \bZ^{(S)}]$, result in aggregations that are sensitive to rescalings or rotations in the underlying latent space of a source ideal point estimate. Moreover, methods like PCA do not explicitly leverage the grouped feature structure (i.e., each source $s$ corresponds to a group of features $\bZ^{(s)}$ of possibly varying dimensions). Consequently, in settings when there are ideal point estimates with varying dimensions, such methods will implicitly construct a consensus ideal point estimate that overweight sources with more dimensions regardless of that source's quality. CoMDS, in contrast, gives equal weight to each source in the objective function \eqref{eqn:optim} and is specifically designed to learn the consensus or shared structure across all input sources. Simulations illustrating these differences are provided in Appendix \ref{app:simulations_nonlinear}-\ref{app:simulations_imbalance}.

Finally, we highlight that CoMDS can naturally accommodate differential amounts of missingness in the underlying data. This allows researchers to input different ideal point estimates that span varying target populations (i.e., legislators, candidates, etc.) and construct a consensus ideal point estimate, even in the presence of nonignorable missing data. CoMDS will embed observations with missing data into a common space alongside complete case observations. See Appendix \ref{app:simulations_missing} for an extended discussion. 

\subsection{Estimating idiosyncratic ideal point components} \label{subsec:idio}
Whereas consensus ideal points enable investigation into substantive relationships between candidates' domain-agnostic ideal points and variables of interest, researchers may also be interested in how different conclusions about accountability, representation, and polarization are specific to certain candidate behaviors \citep[e.g.,][]{meisels_candidate_2025}. Estimating the idiosyncratic components of original source ideal points can shed light on these potential differences across domains. We propose a projection-based approach that decomposes the source ideal point estimates $\bZ^{(s)}$ into two parts: (1) the portion represented by the consensus ideal point estimate $\hat \bZ^{*}$, and (2) an idiosyncratic factor $\hat{\nu}^{(s)}$.

To do so, define the consensus projection matrix $\hat{\mathbf{P}}^* := \bZh^{*} (\bZh^{* \top }\bZh^{*})^{-1} \bZh^{*\top }$. With $\hat{ \mathbf{P}}^*$, we can project the original source ideal point estimates $\bZ^{(s)}$ into the consensus space via $\hat{\mathbf{P}}^* \bZ^{(s)}$. Intuitively, $\hat{\mathbf{P}}^* \bZ^{(s)}$ captures the portion of the $s^{th}$ source ideal point estimate that is represented by the consensus ideal point estimate. We then estimate the idiosyncratic factor for source ideal point $s$ (i.e., $\hat{\nu}^{(s)}$) as the residual term: 
$$\hat \nu^{(s)} := \bZ^{(s)} - \hat{\mathbf{P}}^* \bZ^{(s)} = \left (\mathbf{I} -\hat{\mathbf{P}}^* \right ) \bZ^{(s)},$$
where $\mathbf{I}$ is an $n \times n$ identity matrix. By construction, the estimated idiosyncratic factor $\hat \nu^{(s)}$ will be orthogonal to the consensus ideal point estimate $\hat \bZ^*$. Researchers can equivalently interpret $\hat \nu^{(s)}$ as the residual part of the source ideal point estimate that cannot be explained by the consensus ideal point estimate $\hat \bZ^*$. 

The magnitude of $\hat \nu^{(s)}$ should be interpreted relative to the original, source ideal point estimate scale. For example, NOMINATE ranges from $-1$ to $1$, where values close to $1$ are interpreted as highly conservative, and values close to $-1$ are thought to be highly liberal. In such a setting, when the idiosyncratic factor $\hat \nu^{(s)} > 0$, this implies that relative to the consensus ideal point estimate, NOMINATE has \textit{overestimated} an individual's conservativatism. In contrast, $\hat \nu^{(s)} < 0$ implies that NOMINATE has overestimated an individual's liberalism relative to her consensus ideal point estimate.

\subsection{Diagnostics for CoMDS}\label{subsec:diagnostics}

In the following subsection, we introduce two diagnostics for better understanding each source ideal point measure's contribution to the consensus ideal point estimates. The first is a relative error measure that serves as a normalized proxy for how similar the source ideal points are to the consensus ideal points. Second, we propose a leave-one-out stability analysis to evaluate how consensus ideal point estimates change in response to the omission of different source ideal points.

\paragraph{Relative Error Measure.} To begin, we propose a relative error measure that evaluates the individual contribution of each source ideal point to the consensus ideal point estimate. This is informative of the relative extent to which an existing candidate ideal point measure is idiosyncratic --- in particular, how much of an original source measure remains unexplained by the component which is common across the source measures. Substantively, the relative error diagnostic illuminate whether patterns within a certain domain of candidate activity are substantially orthogonal to the common patterns found across contexts of candidate activity, for instance.

Formally, we define the relative error for a source ideal point estimate $\bZh^{(s)}$ as the squared difference between the pairwise distances in the the $s^{th}$ source ideal point estimate and the consensus ideal point estimate, compared to the total squared difference across all $S$ source ideal points: 
\begin{align*}
    \text{RelError}(\bZ^{(s)}) = 
    \frac{\overbrace{\frac{1}{A^{(s)}} \sum_{i < j} \alpha_i^{(s)} \alpha_j^{(s)}\{ d(\bz_i^{(s)}, \bz_j^{(s)}) - d(\bWh^{(s)} \bzh_i^{*}, \bWh^{(s)} \bzh_j^{*}) \}^2}^{\text{error between } s^{th} \text{ source ideal points and consensus ideal points}} }{\underbrace{\sum_{s = 1}^{S} \frac{1}{A^{(s)}} \sum_{i < j} \alpha_i^{(s)} \alpha_j^{(s)} \{ d(\bz_i^{(s)}, \bz_j^{(s)}) - d(\bWh^{(s)} \bzh_i^{*}, \bWh^{(k)} \bzh_j^{*}) \}^2}_{\text{total error, summed across all sources}}},
\end{align*}
where $A^{(s)} = \sum_{i < j} \alpha_i^{(s)} \alpha_j^{(s)}$.
This source-specific quantity always lies between 0 and 1, where a relative error close to 0 implies that the source ideal point estimate and the consensus ideal point estimate are very similar, whereas a relative error close to 1 implies that the source ideal point estimate and the consensus ideal point estimate are very different. The sum of the relative errors across all $S$ sources equals 1. Consider the extreme setting when the relative error for source $s$ is exactly equal to 1. This implies the relative error for the other sources must be 0. Substantively, this means that source $s$ is orthogonal to all other source ideal point estimates (i.e., there is \textit{no} shared information), while all other sources are identical to one another (up to a rescaling) as well as the consensus ideal point estimate. On the other hand, if the relative error is approximately $\frac{1}{S}$ for each source, then this implies that the consensus ideal point estimate from CoMDS is equally similar to each of the sources.

In practice, scholars rarely have a ``ground truth" measure of an individual's ideal point. As such, a large relative error does not necessarily imply that a source ideal point is necessarily less ``correct" than other sources. However, a large relative error \textit{does} imply that a source ideal point deviates more from the consensus ideal point estimate, and that there is substantial variation in the source ideal point that is not shared by the other source ideal point estimates. For further intuition behind this relative error diagnostic, we refer readers to an illustrative simulation study in Appendix~\ref{app:relative_errors}.

\paragraph{Leave-one-out Stability Analysis.} 
Another way researchers can evaluate the contribution of each source ideal point on the consensus ideal point is through a leave-one-out evaluation. To evaluate the stability of the consensus ideal point estimates to the inclusion or exclusion of different source ideal points, we recommend researchers omit a single data source and re-estimate CoMDS to see if there are large changes in the estimates. We can then evaluate the similarity between the original consensus ideal point estimates $\bZ^*$ and the re-estimated consensus ideal point estimates.

When estimating a one-dimensional ($r = 1$) consensus ideal point, a popular similarity metric is magnitude of the Pearson correlation:
$$\rho(s) = \lvert \cor(\bZh^*_{-(s)}, \bZh^*) \rvert,$$
where $\bZ^*_{-(s)}$ corresponds to the estimated consensus ideal points using CoMDS without including the $s^{th}$ source ideal point estimate. When estimating a multi-dimensional ($r > 1$) consensus ideal point, researchers can instead measure the similarity by computing the subspace correlation between the original consensus ideal point subspace and the re-estimated consensus ideal point subspace when leaving out the $s^{th}$ source. A common measure of subspace correlation is the average squared singular value of the cross-product matrix between the two subspaces \citep{bjorck1973numerical}:
$$\rho(s) = \frac{1}{r}\sum_{i = 1}^{r}d_i^2,$$
where $d_i$ is the $i^{th}$ singular value of $\text{ortho}(\bZh^{*})^{\top}\text{ortho}(\bZh^*_{-(s)})$, and $\text{ortho}(\mathbf{A})$ denotes an orthogonal basis that spans the same subspace as the matrix $\mathbf{A}$. 

At a high level, $\rho(s)$ measures the degree of alignment between $\bZh^{*}$ and $\bZh^*_{-(s)}$, where a larger value close to 1 implies that $\bZh^{*}$ and $\bZh^*_{-(s)}$ are highly similar while a value close to 0 implies that $\bZh^{*}$ and $\bZh^*_{-(s)}$ are orthogonal. In settings when dropping a single source ideal point results in large changes in the consensus ideal points, then $\rho(s)$ will be low. This implies that removing the $s^{th}$ source results in large changes in the resulting consensus ideal point estimates. This occurs in settings when the source ideal point being omitted is very different than the other source ideal points. In contrast, if $\rho(s)$ is high, then this implies that the $s^{th}$ source is similar to the other inputs. As a result, the $s^{th}$ source does not change the consensus ideal point estimate very much. This may occur in settings when multiple source ideal points are similar.

\section{Congressional candidate ideal points} \label{sec:application}

In this section, we apply CoMDS to estimate the ideal points of candidates for the US House. We begin by estimating consensus ideal points based on NOMINATE, CF scores, and campaign platform positions. Despite weak relationships between these three classes of measures, our consensus estimates nevertheless retain substantial relationships with each, including recovering intraparty correlations in cases where one source measure is essentially uncorrelated with others. We then use the consensus measure to assess relationships of substantive interest within domains of the source measures. In particular, the findings demonstrate that relying on existing measures derived from the same source of data as a variable of interest may lead to conclusions which are overstated at best and incorrectly signed at worst. Moreover, we uncover numerous cases in which typical ``robustness checks" --- comparing results across different existing ideal points --- cannot provide satisfactory reconciliation when performed in the absence of a consensus approach. These analyses highlight the broad utility of our consensus ideal points for recovering a common spatial representation in cases where existing ideal points exhibit substantial disagreement, yet researchers would otherwise interpret them interchangeably.

\subsection{Existing measures of House candidate ideal points}

While CoMDS can accommodate virtually any number of measures of congressional candidates' ideal points, here we focus our attention on three classes of existing estimates. Conceptually, these ideal points may represent sincerely--held ideological beliefs, strategic positioning, partisan strength, or a combination therein (e.g. \citealt{lee_beyond_2009, mccarty_defense_2016}). In each case, however, the general quantity of interest is a candidate's position along a left--right continuum. 

The first ideal point, NOMINATE, estimates legislators' ideal points along two dimensions based on a spatial choice model of roll--call voting \citep{lewis_voteview_2025, poole_congress_1997}. Legislators' vote--level decisions are assumed to be solely a function of the distance between their ideal points and the alternative policies represented by Yea versus Nay votes. Given that NOMINATE is based on legislative behavior, it offers coverage for the universe of legislators but necessarily excludes those who fail to win election to Congress. 

Second, we include three flavors of campaign finance (CF) scores \citep{bonica2014mapping, bonica2018rollcall, bonica_database_2024}. The main variant applies correspondence analysis to a matrix of campaign contributions, implicitly assuming that contributions are made on the basis of similarity between donor and recipient. While classic CF scores are static, a second variant is temporally dynamic. The third, DW-DIME, uses machine learning to map contributions onto the NOMINATE space in order to closely predict roll--call behavior. In general, CF scores cover candidates who received contributions from a minimum threshold of contributors who themselves contributed to a minimum threshold of other recipients.

The final measure estimates the positions of issue platforms found on candidates' campaign websites \citep{meisels_candidate_2025}. This dynamic measure relies on the ubiquitous \texttt{wordfish} text scaling algorithm, based on a Poisson IRT model which assumes that word usage is informative of a latent, unidimensional ideal point \citep{slapin_scaling_2008}. Platform positions cover candidates who chose to host a campaign website that included any issue content.

\begin{figure}[t!]
    \centering
    \caption{Relationship Between Source Ideal Point Measures}
    \includegraphics[width=\linewidth]{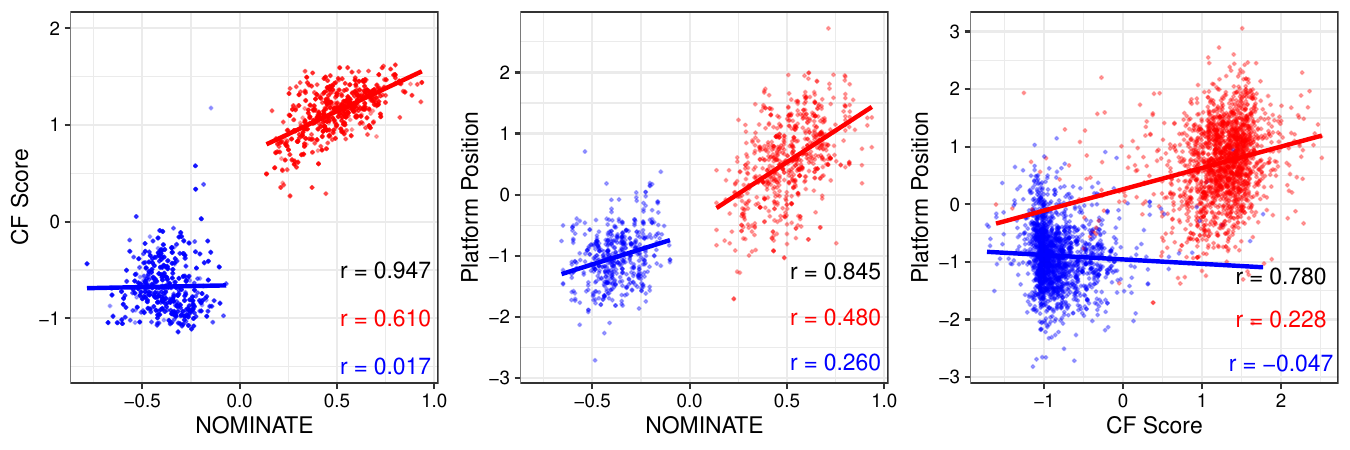}
    \label{fig:sources}
\end{figure}

These existing measures are based on very different sources of data on candidates' activities, which are respectively assumed to be generated by very different processes, and estimation of each relies upon a different statistical approach. For these reasons, it is perhaps unsurprising that past work has shown that different estimates of candidates' ideal points are only weakly related within party \citep{barber_comparing_2022, meisels_candidate_2025, tausanovitch_estimating_2017}. This is confirmed by Figure \ref{fig:sources}: correlations between the three main measures are relatively strong overall yet highly variable within party. Particularly notable is the essentially nonexistent Democratic relationship between CF scores and both of the other measures, echoing results in \citet{barber_comparing_2022} and \citet{meisels_candidate_2025}.\footnote{Complete cases are shown in Appendix \ref{app:robustness}, which reveals modestly stronger correlations between CF scores and platform positions.}

\subsection{Constructing a consensus ideal point estimate}

Despite evidence that these measures are estimated with a considerable amount of domain specificity, they are used and interpreted interchangeably in the vast majority of substantive applications. As articulated straightforwardly by \citeauthor{tausanovitch_estimating_2017}, ``applied empirical studies almost uniformly use the estimates from these models as measures of candidates’ ideology" (\citeyear{tausanovitch_estimating_2017}, 168). Because researchers studying topics such as polarization, representation, or accountability are typically interested in characterizing candidates' general, domain--agnostic extremism versus moderation or liberalism versus conservatism, contextual differences between measures are downplayed and typically treated as little more than a nuisance.\footnote{To be clear, we concur with calls for scholars to carefully consider the applicability of their theories to different domains. Our projection-based approach to estimating the remaining idiosyncratic components of existing measures is particularly well-suited for future work on substantive differences between candidates' ideal points across contexts.}

Our consensus-based approach provides a closer approximation of how existing estimates of candidates' ideal points tend to be used in practice. We estimate the consensus ideal points of House candidates from 2016 to 2024 using the roll--call--based, contribution--based, and platform--based measures discussed above. Since CoMDS can handle missing data without relying on imputation, we include all candidates captured by at least two of the three measures. In effect, we are able to scale non--incumbents who have both platform positions and CF scores as well as all incumbents, leaving us with a total of 5,389 unique candidate--year observations.\footnote{We provide a stability analysis of the estimated consensus ideal points in Appendix \ref{sec:pcs}, where we show that the estimated consensus ideal points are robust to alternative analysis choices, such as using just the first dimension of NOMINATE, or using only complete data.}

\begin{figure}[t!]
    \centering
    \caption{Distribution of Candidates' Consensus Ideal Points by Party}
    \includegraphics[width=.7\linewidth]{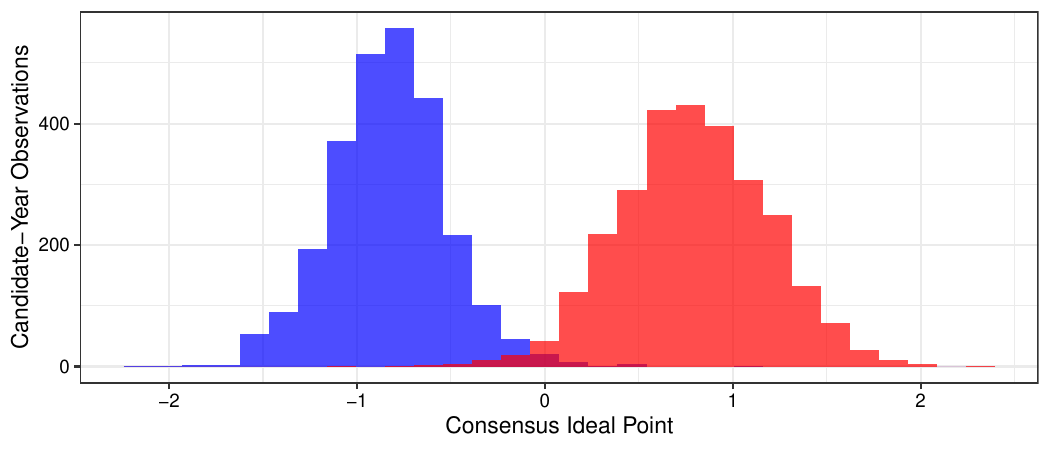}
    \label{fig:histogram}
\end{figure}

\begin{figure}[t!]
    \centering
    \caption{Relationship Between Source and Consensus Ideal Point Measures}\label{fig:compare}
    \includegraphics[width=\linewidth]{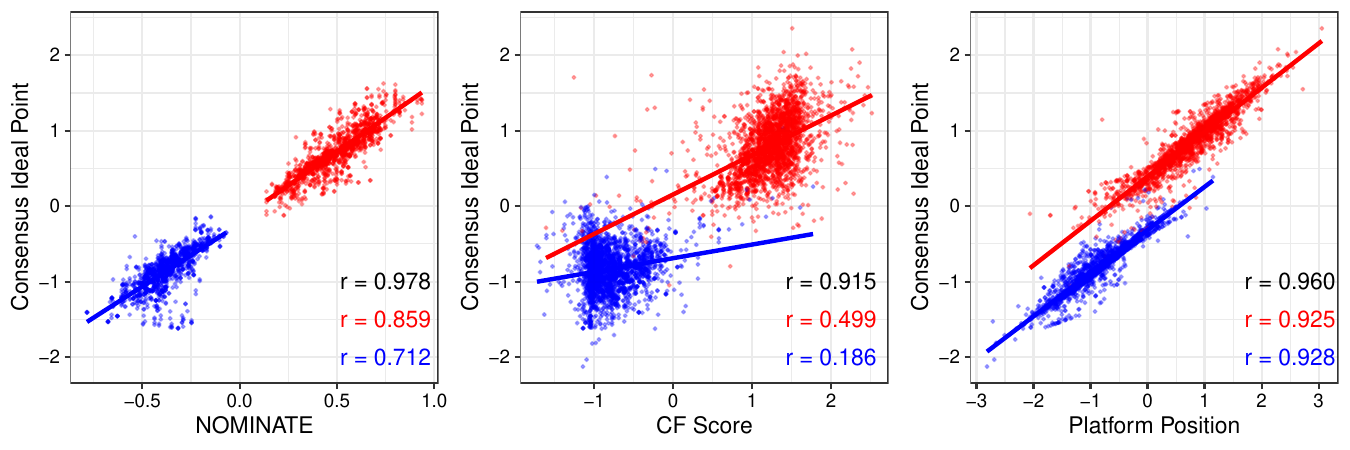}
\end{figure}

The distribution of consensus ideal points in Figure \ref{fig:histogram} demonstrates an expectedly strong partisan bimodality, with the vast majority of Democrats falling to the left of the vast majority of Republicans.\footnote{In Appendix \ref{app:robustness}, we report and discuss relationships between ideal points estimated via CoMDS versus the most similar alternative aggregation approaches (MD2S and PCA).} Moreover, Figure \ref{fig:compare} suggests that the consensus ideal points maintain modestly strong correlations between each of the main source measures. In addition to characteristically high correlations overall, each original measure exhibits stronger relationships with the consensus measure than with the other measures as reported in Figure \ref{fig:sources}. For example, despite the fact that CF scores are essentially uncorrelated with the other source measures among Democrats --- the correlation with platforms is slightly negative while correlation with NOMINATE is 0.017 --- Democrats' CF scores correlate with their consensus ideal point at nearly 0.2.

\subsection{Assessing substantive relationships within domains of existing measures}

As discussed above, existing estimates of candidate ideal points tend to be used and interpreted interchangeably in applied empirical research despite evidence suggesting that they are measured with a considerable amount of domain specificity. This issue is exacerbated by the fact that many research questions pertaining to elections, representation, polarization, and accountability implicate a variable of interest derived from the same source of data as an existing ideal point. For example, consider the question of whether extremism is beneficial or detrimental for fundraising in House elections. In its simplest form, investigation necessitates some transformation of campaign contribution data on the left--hand side of a regression equation, and some transformation of candidate ideal points --- potentially including CF scores --- on the right--hand side.

In the absence of a consensus approach, researchers without \textit{a priori} expectations about differences across measures must resort to comparing the ``robustness" of results obtained using different domain--specific measures, including one within the domain of the substantive variable of interest. If a strong relationship between an ideal point and a variable of interest in the same domain is uncovered, this may be substantively meaningful or merely a mechanical result of the variables being measured more-or-less jointly. Any disagreement about statistical or substantive significance across domain--specific ideal points will therefore pose a potentially prohibitive obstacle to drawing conclusions about relationships of interest. By identifying the shared, stable associations between different measures, our consensus approach distances ideal points from any one particular domain, facilitating a sounder assessment of how variables within domains of existing ideal points relate to a more domain--agnostic ideal point.\footnote{In Appendix \ref{app:robustness}, we report all results based solely on complete cases (i.e., legislators with campaign website platforms) or including only the three main measures in CoMDS estimation (i.e., classic CF scores, platform positions, and first-dimension NOMINATE). We also report and discuss results using an existing consensus ideal point measure --- composite scores from \citet{bonica_database_2024} --- although these are estimated based on both different source measures and aggregation approaches.}

\subsubsection{Ideal points and roll--call partisan disloyalty}

The connection between roll--call ideal points and partisan loyalty in roll--call voting among members of Congress is well--established, with extreme legislators voting in lockstep with their parties at higher rates than moderates \citep{carson_electoral_2010,minozzi_who_2013}. However, this may be partly a function of the estimation of roll--call ideal points such as NOMINATE. Because members who vote together are implicitly assumed to have more similar ideal points, those who frequently vote against their party --- likely joining the opposing party in doing so --- will have roll--call ideal points closer to the opposing party than their co-partisans who never break with the party. In fact, this is precisely the culprit of NOMINATE's ``AOC problem", wherein NOMINATE places known progressives squarely in the moderate wing of the Democratic caucus \citep{lewis_voteview_nodate}. Such models fail to account for ``ends against the middle" behavior, i.e., dissimilar members opposing legislation for opposite reasons \citep{duck-mayr_ends_2023}.\footnote{The canonical cases for ``The Squad" involve voting against Democratic legislation they deemed insufficiently progressive, joining members of the Republican Party who deemed the same legislation insufficiently conservative.}

We re-examine the relationship between legislators' ideal points and party loyalty by performing two sets of analyses. First, we compare results based on consensus ideal points versus NOMINATE to assess how relying on an ideal point within the same domain as the dependent variable may distort conclusions. Second, we compare results across the three main source measures --- NOMINATE, CF scores, and platform positions. In the absence of a consensus approach, researchers would perform such a ``robustness check" to attempt to rule out whether a result is simply an artifact of independent and dependent variables being derived from the same domain. We maximize comparability by standardizing all ideal points within--sample and including only observations covered by the relevant measures being compared in each analysis. Consequently, any differences in results will be due to differences in the ideal point estimates rather than differences in sample coverage.

Our dependent variable, partisan disloyalty in roll--call voting, is based on Congressional Quarterly's (CQ) longstanding measure of party unity. CQ identifies party unity roll--call votes as those where the majority of one party voted in opposition to the majority of the other party, and legislators' party unity scores are the share of party unity votes on which the legislator voted with her party. Unsurprisingly, these scores are heavily left--skewed: few members side with their party on fewer than 90\% of party unity votes. We therefore calculate partisan disloyalty --- a more informative measure --- by log-transforming party unity subtracted from one.\footnote{Plotting ideal points against all dependent variables of interest in Appendix \ref{app:robustness} suggests strong nonlinearities in the relationships, so we also control for a quadratic specification of respective ideal point measures in all reported models.}

Estimates in the first two columns of Table 1 suggest that domain--agnostic and roll--call--specific ideal points are in agreement about the directionality of the relationship between legislators' moderation and roll--call partisan disloyalty. As Democrats' consensus ideal points grow more conservative/less liberal, they vote against their party significantly more often, whereas Republicans vote against their party significantly less often as their consensus ideal points become more conservative/less liberal. The magnitude of the coefficients from the NOMINATE models are exaggerated relative to the consensus models, however, suggesting that measuring independent and dependent variables with the same roll--call data inflates the size of relationships.

The consensus ideal points allow us to conclude that generally moderate legislators vote against their party much more frequently than generally extreme legislators. However, comparing estimates across source measures in Table 1 suggests that without our consensus approach, it would be difficult to ascertain this relationship when checking for robustness across existing ideal points. Among Democrats, no relationship is detected between CF scores and partisan disloyalty. The estimated magnitude of the relationship based on platform positions is less than one third that of NOMINATE. Relationships between the source ideal points and partisan disloyalty are more consistent among Republicans, but their magnitudes vary considerably across measures, despite the fact that models only include legislators covered by all three. Taken together, these results elucidate how the consensus approach allows us to draw meaningful conclusions about the strength and size of the relationship between legislators' moderation and partisan disloyalty in roll--call voting when component measures produce varied estimates.

\begin{table}
\centering\centering
\caption{Relationship Between Ideal Points and Roll-Call Partisan Disloyalty}
\centering
\fontsize{10}{12}\selectfont
\begin{tabular}[t]{lccccc}
\toprule
\multicolumn{1}{c}{ } & \multicolumn{5}{c}{DV: log(\% Votes Opposing Party Majority)} \\
\cmidrule(l{3pt}r{3pt}){2-6}
\multicolumn{1}{c}{ } & \multicolumn{2}{c}{Consensus vs. Domain} & \multicolumn{3}{c}{Comparing Source Measures} \\
\cmidrule(l{3pt}r{3pt}){2-3} \cmidrule(l{3pt}r{3pt}){4-6}
  & Consensus & NOMINATE & NOMINATE  & CF Score & Platform\\
\midrule
\addlinespace[0.5em]
\multicolumn{6}{l}{\textit{Panel A. Democrats}}\\
\midrule \hspace{1em}Ideal Point & 0.673*** & 0.727*** & 0.729*** & 0.036 & 0.211***\\
\hspace{1em} & (0.035) & (0.027) & (0.043) & (0.073) & (0.062)\\
\hspace{1em}Observations & 1,206 & 1,206 & 493 & 493 & 493\\
\midrule
\addlinespace[0.5em]
\multicolumn{6}{l}{\textit{Panel B. Republicans}}\\
\midrule \hspace{1em}Ideal Point & -0.504*** & -0.651*** & -0.628*** & -0.352*** & -0.291***\\
\hspace{1em} & (0.034) & (0.028) & (0.040) & (0.048) & (0.047)\\
\hspace{1em}Observations & 1,340 & 1,340 & 660 & 660 & 660\\
\bottomrule
\end{tabular}
\\\footnotesize \textit{Note:} Ideal points are increasing in conservatism and rescaled within sample to have mean 0 and SD 1. Models control for second-order polynomial. Legislator–clustered standard errors in parentheses.* p $<$ 0.05, ** p $<$ 0.01, *** p $<$ 0.001
\end{table}

\subsubsection{Ideal points and campaign contributors}

Whether extremists are financially advantaged in elections is a central question in recent literature on congressional polarization. Individual donors are demonstrably extreme compared to other members of the population, including voters, and much work suggests that they contribute to extreme candidates on the basis of shared positions \citep{ansolabehere_why_2003, barber_representing_2016, kujala_donors_2020}. Other scholarship has shown, however, that individual donors are likewise motivated by more strategic considerations which may also lead them to support non--extremists \citep{gimpel_check_2008, meisels_giving_2024}. In contrast to individuals, evidence suggests that political organizations --- and especially corporations --- tend to support moderates \citep{barber_ideological_2016, meisels_everything_2025, thieme_moderation_2020}. Reliance on contribution--based ideal points is likely problematic for examining the relationship between candidate positioning and fundraising success as candidates who are high--profile (such as incumbents) are in a position to raise funds from nationalized donors, whereas the vast majority of lower--profile candidates will necessarily rely on local support. 

Similarly to the previous section, we perform two sets of analyses to re-examine the relationship between candidates' ideal points and bases of financial support. The first compares results based on consensus ideal points versus domain--specific CF scores, while the second investigates how ``robust" results would appear when comparing across existing ideal point measures. To operationalize financial support, we log--transform the number of distinct contributors who gave to a candidate over the course of her career as reported in the Database on Ideology, Money, and Elections \citep{bonica_database_2024}. Focusing on number of donors rather than donation totals helps to avoid capturing differences in the wealth of donors who support candidates with different ideal points, instead capturing differences in the general size of candidates' bases of support. Additionally, we take the same steps as before to maximize comparability across ideal point models by rescaling measures and subsetting to common observations.

Estimates from the first two columns of Table 2 suggest that basic conclusions about candidate extremism and financial support differ substantially depending upon whether consensus or domain--specific ideal points are used. Among both Democrats and Republicans, financial base decreases substantially with more conservative consensus ideal points. This effectively suggests that general extremism is financially advantageous for Democrats whereas general moderation is financially advantageous for Republicans. In contrast, there is a positive association between CF score moderation and financial support among candidates of both parties, and coefficient magnitudes are more than twice as large in CF score models than in consensus ideal point models. Therefore, the shared component across existing ideal point measures --- as captured by consensus ideal points --- exhibits a fundamentally different relationship with candidates' financial support than CF scores, both in terms of substantive size and even directionality. 

\begin{table}
\centering\centering
\caption{Relationship Between Ideal Points and Fundraising Success}
\centering
\fontsize{10}{12}\selectfont
\begin{tabular}[t]{lccccc}
\toprule
\multicolumn{1}{c}{ } & \multicolumn{5}{c}{DV: log(Number of Unique Campaign Donors)} \\
\cmidrule(l{3pt}r{3pt}){2-6}
\multicolumn{1}{c}{ } & \multicolumn{2}{c}{Consensus vs. Domain} & \multicolumn{3}{c}{Comparing Source Measures} \\
\cmidrule(l{3pt}r{3pt}){2-3} \cmidrule(l{3pt}r{3pt}){4-6}
  & Consensus & CF Score & NOMINATE & CF Score  & Platform\\
\midrule
\addlinespace[0.5em]
\multicolumn{6}{l}{\textit{Panel A. Democrats}}\\
\midrule \hspace{1em}Ideal Point & -0.373*** & 0.839*** & 0.221** & -0.811*** & 0.051\\
\hspace{1em} & (0.073) & (0.072) & (0.073) & (0.065) & (0.066)\\
\hspace{1em}Observations & 2,629 & 2,629 & 516 & 516 & 516\\
\midrule
\addlinespace[0.5em]
\multicolumn{6}{l}{\textit{Panel B. Republicans}}\\
\midrule \hspace{1em}Ideal Point & -0.423*** & -0.963*** & -0.111* & 0.350*** & -0.120*\\
\hspace{1em} & (0.062) & (0.057) & (0.055) & (0.066) & (0.061)\\
\hspace{1em}Observations & 2,760 & 2,760 & 678 & 678 & 678\\
\bottomrule
\end{tabular}
\\\footnotesize \textit{Note:} Ideal points are increasing in conservatism and rescaled within sample to have mean 0 and SD 1. Models control for second-order polynomial. Candidate–clustered standard errors in parentheses.* p $<$ 0.05, ** p $<$ 0.01, *** p $<$ 0.001
\end{table}

Given the discrepancy between results based on domain--specific versus domain--agnostic measures, it is perhaps unsurprising that a ``robustness check" comparing results across different existing ideal points in the absence of our consensus approach does not provide a clear substantive takeaway. Note that estimates from CF score models in Table 3 are of different signs when comparing to the consensus model versus other source measure models. While the former includes all candidates covered by CF scores and either NOMINATE or platform positions, the latter includes only candidates covered by all three. As noted by \citet{meisels_candidate_2025}, non-incumbents (who tend to be worse fundraisers) have far more extreme CF scores than incumbents, yet this is not the case for other measures. Even among the complete cases covered by all three measures, there is substantial disagreement across source measures. While CF scores suggest that extremism is associated with a greater financial base, platform positions and NOMINATE suggest that, if anything, moderation is associated with a greater financial base. In all, the comparison of results across existing ideal points demonstrates that absent a consensus approach such as ours, researchers would be precluded from drawing a coherent overall conclusion.

\subsubsection{Ideal points and lexical diversity}

While much work investigating speeches, press releases, and platforms has focused on ideological content, complexity is another important dimension of political texts. The ``dumbing down" of political rhetoric across time, particularly with regard to State of the Union addresses, has received both popular and scholarly attention \citep{benoit_measuring_2019}. Related research has investigated how electorate expansion as manifested via election rounds \citep{di_tella_keep_2025} or franchise extension \citep{spirling_democratization_2016} corresponds to the linguistic sophistication of elite rhetoric. Another strand of work evaluates the relationship between ideological positions and linguistic complexity of candidates/parties (e.g. \citealt{schoonvelde_liberals_2019, tetlock_cognitive_1983}), generally finding that greater conservatism is associated with less complex language. These studies typically compare the rhetorical sophistication of politicians belonging to conservative and liberal parties (e.g. Republicans and Democrats in the US).

To assess whether there exists a \textit{within-party} relationship between conservatism and lexical diversity, we once again perform analyses comparing results based on consensus ideal points to results based on domain--specific ideal points, as well as comparing results across three existing ideal point measures. To capture lexical diversity, we rely on campaign website platforms --- one of the only longform corpuses covering large swaths of non-incumbents \citep{meisels_candidate_2025}.\footnote{In contrast, numerous sources of data can be used to study sitting legislators' lexical diversity, such as press releases or congressional speeches.} The aforementioned research on linguistic sophistication employ a wide variety of measures which consider features such as the length and complexity of words, sentences, and documents as a whole. Unlike the more formal texts of interest in such studies, campaign platforms frequently feature bulleted lists and many subheadings which are not complete sentences. Because the syntax of this corpus does not lend itself well to metrics relying upon the syntactical features of natural sentences, we rely on a simple metric of lexical diversity, Carroll's Corrected Type-Token Ratio (CTTR) \citep{carroll_language_1964}. This measure proxies conceptual complexity of text using the ratio of unique words to total words in the document, then applies a correction to account for the ratio's sensitivity to document length by swapping out the total words in the denominator for the square root of double the total words.

\begin{table}
\centering\centering
\caption{Relationship Between Ideal Points and Lexical Diversity}
\centering
\fontsize{10}{12}\selectfont
\begin{tabular}[t]{lccccc}
\toprule
\multicolumn{1}{c}{ } & \multicolumn{5}{c}{DV: Campaign Platform Corrected Type-Token Ratio (CTTR)} \\
\cmidrule(l{3pt}r{3pt}){2-6}
\multicolumn{1}{c}{ } & \multicolumn{2}{c}{Consensus vs. Domain} & \multicolumn{3}{c}{Comparing Source Measures} \\
\cmidrule(l{3pt}r{3pt}){2-3} \cmidrule(l{3pt}r{3pt}){4-6}
  & Consensus & Platform & NOMINATE & CF Score & Platform \\
\midrule
\addlinespace[0.5em]
\multicolumn{6}{l}{\textit{Panel A. Democrats}}\\
\midrule \hspace{1em}Ideal Point & 0.376*** & 0.471*** & 0.156 & -0.344 & 0.287\\
\hspace{1em} & (0.090) & (0.080) & (0.193) & (0.203) & (0.178)\\
\hspace{1em}Observations & 1,836 & 1,836 & 516 & 516 & 516\\
\midrule
\addlinespace[0.5em]
\multicolumn{6}{l}{\textit{Panel B. Republicans}}\\
\midrule \hspace{1em}Ideal Point & -0.542*** & -0.729*** & -0.064 & 0.057 & -0.633***\\
\hspace{1em} & (0.064) & (0.064) & (0.122) & (0.144) & (0.129)\\
\hspace{1em}Observations & 1,994 & 1,994 & 678 & 678 & 678\\
\bottomrule
\end{tabular}
\\\footnotesize \textit{Note:} Ideal points are increasing in conservatism and rescaled within sample to have mean 0 and SD 1. Models control for second-order polynomial. Candidate–clustered standard errors in parentheses.* p $<$ 0.05, ** p $<$ 0.01, *** p $<$ 0.001
\end{table}

Results using both consensus and domain--specific ideal points in Table 3 demonstrate that intraparty moderation --- not liberalism, as past studies suggest --- is associated with greater rhetorical sophistication. However, among both Democrats and Republicans, consensus--based estimates are more modest than platform--based estimates. Comparing coefficients across source measures, however, suggests that this relationship would be undetectable when performing a robustness check in the absence of our consensus approach. While the platform--based estimates are smaller (and in the case of Democrats, statistically insignificant at traditional levels) among complete cases compared to incomplete cases, the other two ideal point measures disagree on sign and magnitude. Without the estimates based on consensus ideal points, it would be unclear whether strong relationships solely in the case of platform--based ideal points are simply due to being measured using the same source of data as the dependent variable.

\section{Conclusion} 

Across political science subfields, numerous ongoing debates require estimates of political actors' ideal points. Unsettled research questions at the forefront of American politics, for example, necessitate estimates of congressional candidates' positions. Extant work documents weak intraparty relationships between estimates based on different sources of data and measurement approaches, suggesting that each measure may be a mixture of a more domain--agnostic ideal point common across measures, as well as a domain--specific, idiosyncratic component \citep{barber_comparing_2022, meisels_candidate_2025, tausanovitch_estimating_2017}. However, the quantity of interest in most applied studies is a more general concept of candidate ``positioning" or ``ideology", leading scholars to treat different measures more-or-less interchangeably. 

We propose consensus multidimensional scaling (CoMDS) as an estimation approach which better aligns with how ideal points tend to be used in practice. While assessment of results' robustness across measures is commonly employed to ensure findings are not driven by any one ideal point measure, we have shown that existing measures may disagree about not just the substantive size of basic relationships at the forefront of ongoing debates, but also their directionality. Therefore, in the absence of a consensus approach such as ours, researchers will be precluded from drawing meaningful conclusions about relationships of interest. In contrast, consensus ideal points facilitate investigation of relationships between the component which is \textit{shared} across alternative ideal point measures and variables of interest.

Beyond the usage of consensus ideal points of congressional candidates to investigate additional relationships of interest, there are a number of aspects of the approach introduced here which open up avenues of future research. CoMDS does not require researchers to jointly model the outcomes from different data sources, allowing for encoding of different underlying behavioral models across contexts. This flexibility is further enhanced by the ability to handle differential amounts of missingness across source measures, and resulting consensus ideal points are likewise invariant to rescalings or rotations of source measures.

The customizability of CoMDS makes it well--suited for the estimation of other actors' ideal points as well as the incorporation of additional source measures as they emerge. While we have focused on three broad classes of existing congressional candidate ideal points, newer estimates from LLMs or social media venues can be easily included alongside other source measures. This allows for future examination of whether and how consensus ideal points change with the inclusion of newer estimates. Beyond the specific context of focus in our application, CoMDS may be useful in other settings where there exist multiple measures of similar concepts which exhibit substantial disagreement in practice. Because there is rarely a ``ground truth" for latent ideal points almost by definition, our consensus approach offers a data--driven, reconciliatory approach for researchers who are uninterested in contextual differences between alternative measures of similar concepts. 

On the flip side of identifying their shared component, the accompanying projection--based approach we propose facilitates further investigation into the \textit{differences} between source measures. While there is clear evidence on the relatively weak relationships between estimates of congressional candidates' ideal points, it has been much more difficult to pinpoint precisely why they differ. By decomposing each measure into a component shared with the other measures and a remaining idiosyncratic, orthogonal component, future work can more easily analyze relationships between the domain--specific idiosyncratic factors and external variables of interest. In turn, this can spur new avenues of research into differences in candidates' strategic behavior across venues and activities.

\clearpage
\singlespacing
\bibliographystyle{chicago}
\bibliography{bibliography_arxiv}

\clearpage 

\appendix
\setcounter{page}{1}

\appendix
\setcounter{page}{1}
\setcounter{figure}{0}
\setcounter{table}{0}

\renewcommand{\partname}{}
\renewcommand{\thepart}{}

\part{Appendix}

\localtableofcontents

\newpage

\section{Additional Discussion} 
\subsection{Relationship with MD2S \citep{enamoradoScalingDataMultiple2021}}
\label{app:ted}

\setcounter{figure}{0}
\setcounter{table}{0}
\renewcommand{\thefigure}{A\arabic{figure}}
\renewcommand{\thetable}{A\arabic{table}}

Closest related to our method is an approach introduced in \citet{enamoradoScalingDataMultiple2021} called MD2S. \citet{enamoradoScalingDataMultiple2021} focus on the setting in which researchers have two data sources and have access to the outcomes from both data sources: $Y^{(1)}, Y^{(2)} \in \R^{n \times p}$. Then, they assume the following linear data generating process:
\begin{align} 
\begin{cases} 
Y^{(1)} &= \bZ^{*} \boldsymbol{\Gamma}^{(1)} + \nu^{(1)} + \varepsilon^{(1)}\\
Y^{(2)} &= \bZ^{*} \boldsymbol{\Gamma}^{(2)} + \nu^{(2)}+ \varepsilon^{(2)}
\end{cases},
\label{eqn:m2ds}
\end{align} 
where to estimate the model, they further impose low-rank assumptions on $\nu^{(s)}$, orthogonality constraints on $\boldsymbol{\Gamma}^{(s)}$ and $\nu^{(s)}$, and normality assumptions on $\varepsilon^{(s)}$. The proposed algorithm then estimates $\{\bZ^*, \boldsymbol{\Gamma}^{(1)}, \boldsymbol{\Gamma}^{(2)}, \nu^{(1)}, \nu^{(2)}\}$ by jointly modeling both $Y^{(1)}$ and $Y^{(2)}$.

Comparing the assumed data generating process in Equation \eqref{eqn:m2ds} with the assumed data generating process for CoMDS in Equation \eqref{eqn:idealpoint_model} helps highlight the differences in the methods. Recall, in Equation \eqref{eqn:idealpoint_model}, we assume that the outcomes can be written as $Y^{(s)} = g^{(s)}(\bZ^* + \nu^{(s)}) + \varepsilon^{(s)}$. However, we do not place any restrictions on the functional form of $g^{(s)}$ or the error term $\varepsilon^{(s)}$. As such, we can view the data generating process in Equation \eqref{eqn:m2ds} as a special case of Equation \eqref{eqn:idealpoint_model}, where the function $g^{(s)}$ is equal to a \textit{linear} mapping for all data sources. Assuming that all outcomes must have a linear relationship with the true ideal point $\bZ^*$ can be relatively restrictive and does not map to the underlying behavioral models posited for many existing data sources. For example, NOMINATE often assumes that yea/nay votes follow a logistic function of the underlying ideal points. 

\subsection{Interpreting CoMDS as a constrained MDS problem} \label{app:mds}
Another way to interpret CoMDS is by considering its equivalency as a constrained multidimensional scaling problem. More specifically, define the block diagonal matrix $\bD = diag(\bD^{(1)}, \ldots, \bD^{(S)}) \in \mathbb{R}^{Sn \times Sn}$, and suppose we applied ordinary multidimensional scaling (MDS) \citep{Torgerson-MDS} to $\bD$ with a particular constraint --- specifically,
\begin{align}
    \hat{\bZ}^{*}_{MDS} = \argmin_{\tilde{\bZ} \in \R^{Sn \times r}} \sum_{i < j} \left( \bD_{ij} - d(\tilde{\bz}_i, \tilde{\bz}_j) \right)^2,\label{eqn:constrained_mds}
\end{align}
subject to the following constraints: 
\begin{align*}
    (i)~~\tilde{\bZ} = \begin{bmatrix} \bZ \bW^{(1)} \\ \vdots \\ \bZ \bW^{(S)} \end{bmatrix},\; (ii)~ \bW^{(1)}, \ldots, \bW^{(S)} &\text{ are } r \times r \text{ diagonal matrices, and } 
    (iii)~\bZ \in \mathbb{R}^{n \times r}.
\end{align*}
Then the constrained MDS solution \eqref{eqn:constrained_mds} satisfies
\begin{align}
    \hat{\bZ}^{*}_{MDS} = \begin{bmatrix} \hat{\bZ}^{*} \hat{\bW}^{(1)} \\ \vdots \\ \hat{\bZ}^{*} \hat{\bW}^{(S)} \end{bmatrix},
\end{align}
where $\hat \bZ^*$ and $\hat \bW^{(1)}, \ldots, \hat \bW^{(S)}$ are the solution to the CoMDS problem from \eqref{eqn:optim}.

\section{Illustrative simulations} \label{app:simulations}

\setcounter{figure}{0}
\setcounter{table}{0}
\renewcommand{\thefigure}{B\arabic{figure}}
\renewcommand{\thetable}{B\arabic{table}}

In this section, we provide a series of illustrative simulations to build up intuition behind CoMDS and to highlight key advantages of CoMDS over existing approaches. In particular, we compare CoMDS to simply applying PCA to the concatenated source ideal points $[\bZ^{(1)}, \ldots, \bZ^{(S)}]$, concretely demonstrating four important advantages of CoMDS --- namely, (i) its flexibility for handling nonlinear data-generating processes in Section~\ref{app:simulations_nonlinear}, (ii) its invariance to arbitrary shifts, rescalings, and rotations in the source ideal points in Section~\ref{app:simulations_rotation}, (iii) its ability to identify consensus patterns even under imbalanced and correlated settings in Section~\ref{app:simulations_imbalance}, and (iv) its inherent ability to handle missing data.

\subsection{Nonlinear data-generating simulation}\label{app:simulations_nonlinear}

To begin, recall the posited ideal point model \eqref{eqn:idealpoint_model} from Section~\ref{sec:ideal_points}, where we assume the data source $Y^{(s)}$ is generated from some function $g_s(\cdot)$ of the ideal point estimates $\bZ^{(s)}$ plus possible noise $\epsilon^{(s)}$:
$$Y^{(s)} = g_s(\bZ^{(s)}) + \epsilon^{(s)}.$$
Previous methods such as PCA \citep{bonica_database_2024} and MD2S \citep{enamoradoScalingDataMultiple2021}\footnote{The matrix factorization model of MD2S is equivalent to that of JIVE \citep{lock2013joint}, a popular data integration method from the biomedical literature. While MD2S was only introduced for integrating two datasets, JIVE provides the general formulation and software for integrating any arbitrary number of datasets. We hence use the JIVE software implementation for all empirical comparisons in this work.} assume that $g_s(\cdot)$ takes the form of a linear function. This linear assumption is often restrictive and may not hold in practice. CoMDS, on the other hand, does not assume a linear form for $g_s(\cdot)$ and instead aims to preserve the pairwise distances between samples in the source ideal point space and in the consensus ideal point space. This distance-preserving property gives CoMDS more flexibility to capture possibly nonlinear relationships in $g_s(\cdot)$.

\begin{figure}[tb]
    \centering
    \includegraphics[width=0.9\linewidth]{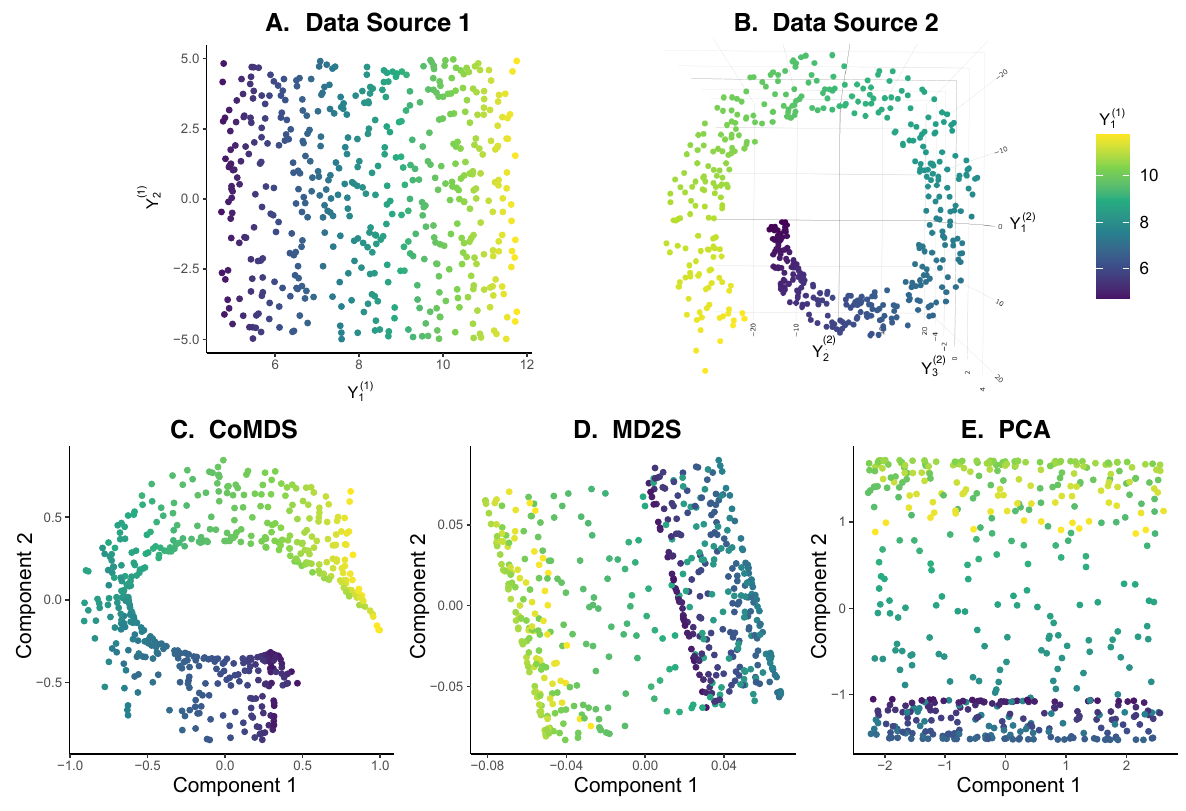}
    \caption{(A) We simulate one data source $Y^{(1)} \in \R^{500 \times 2}$ from a uniform distribution. (B) We then apply the nonlinear swiss roll transform to $Y^{(1)}$ to obtain a second data source $Y^{(2)} \in \R^{500 \times 3}$. (C-E) The consensus plots from applying CoMDS, MD2S, and PCA to $Y^{(1)}$ and $Y^{(2)}$ are respectively shown. Linear-based approaches such as MD2S and PCA ``flatten'' the swiss roll structure and lose the original ordering of the ideal points. In contrast, CoMDS ``unravels'' the swiss roll and more accurately preserves the original ideal point structure.}
    \label{fig:swiss_roll}
\end{figure}

To illustrate this ability to handle nonlinearity, we simulate two data sources $Y^{(1)}$ and $Y^{(2)}$ using the same underlying set of 2-dimensional ideal point estimates $\bZ^{*}$ --- one with a linear $g_s(\cdot)$ and one with a nonlinear $g_s(\cdot)$. Specifically, we simulate the true consensus ideal point estimates $\bZ^{*}$ from a two-dimensional uniform distribution with $n = 500$ samples. We then let the first data source $Y^{(1)} = \bZ^{*} \in \R^{500 \times 2}$ (i.e., $g_s(\cdot)$ is the identity function) and the second data source $Y^{(2)} \in \R^{500 \times 3}$ to be the nonlinear ``swiss roll'' transformation of $\bZ^{*}$ \citep{tenenbaum2000global}\footnote{The swiss roll is a popular toy nonlinear transformation, wherein a 2-dimensional flat surface is ``rolled'' up like a Swiss roll pastry in 3-dimensional space.}. These two simulated data sources are depicted in Figure~\ref{fig:swiss_roll}A-B.

Since both data sources $Y^{(1)}$ and $Y^{(2)}$ are direct (noise-less) functions of $\bZ^{*}$, we would expect that the estimated consensus ideal points $\bZh^{*}$ resemble $\bZ^{*}$ (which is equivalent to $Y^{(1)}$ shown in Figure~\ref{fig:swiss_roll}A). However, as shown in Figure~\ref{fig:swiss_roll}D-E, applying linear-based approaches such as MD2S and PCA do not preserve the original ordering of ideal points (from yellow to dark purple). MD2S and PCA are both designed to find a linear projection of the data sources and thus must ``flatten'' the swiss roll structure, resulting in the outer tails of the roll (i.e., the bright yellows and the dark purples) being compressed towards the center of the consensus space rather than being preserved as the outer edges of the swiss roll. In contrast, CoMDS does not assume linearity and instead focuses on preserving pairwise distances. This results in the estimated CoMDS consensus plot shown in Figure~\ref{fig:swiss_roll}C, which preserves the nice gradient from yellow to dark purple.

\subsection{Rotation invariance simulation}\label{app:simulations_rotation}

Another key advantage of CoMDS is its robustness and invariance to shifts, rescalings, and rotations of the source ideal points. 
Because the numerical values of ideal points are in essence arbitrary and it's only the relative distances between ideal points that are readily interpretable, being invariant to such shifts, rescalings, and rotations is a desirable property. Put differently, it is desirable that the estimated consensus ideal points do not change if the source ideal points are arbitrarily shifted, rescaled, or rotated in any way.\footnote{Importantly, these transformations do not change the meaning of the source ideal points.} However, methods such as PCA, which operate on the raw numerical values of the source ideal points, are sensitive to such transformations and can yield different results depending on the orientation or scaling of the input embeddings.

To illustrate the invariance property of CoMDS, we will specifically focus on the rotation invariance and construct a toy simulation, wherein we first simulate the true underlying consensus ideal points $\bZ^{*} \in \R^{100 \times 2}$ with independent and identically distributed entries from a standard normal distribution. We then consider two hypothetical scenarios:
\begin{enumerate}
    \item \textit{Original Scenario:} Applying CoMDS or PCA to estimate the consensus ideal points from the two (unrotated) data sources: $Y^{(1)} = \bZ^{*}$ and $Y^{(2)} = \bZ^{*}$;
    \item \textit{Rotated Scenario:} Applying CoMDS or PCA to estimate the consensus ideal points from the two data sources after rotating one by 60 degrees: $Y^{(1)} = \bZ^{*}$ and $Y^{(2)} = \bZ^{*} \mathbf{R}$, where $\mathbf{R}$ is a 2-dimensional rotation matrix that rotates the ideal points by 60 degrees.
\end{enumerate}
Since the substantive meaning and interpretation of $\bZ^{*} \mathbf{R}$ is the same as that of the unrotated $\bZ^{*}$, we would like for the estimated consensus ideal points $\bZh^{*}$ from the two hypothetical scenarios to be equivalent to each other. 

\begin{figure}[tb]
    \centering
    \includegraphics[width=0.95\linewidth]{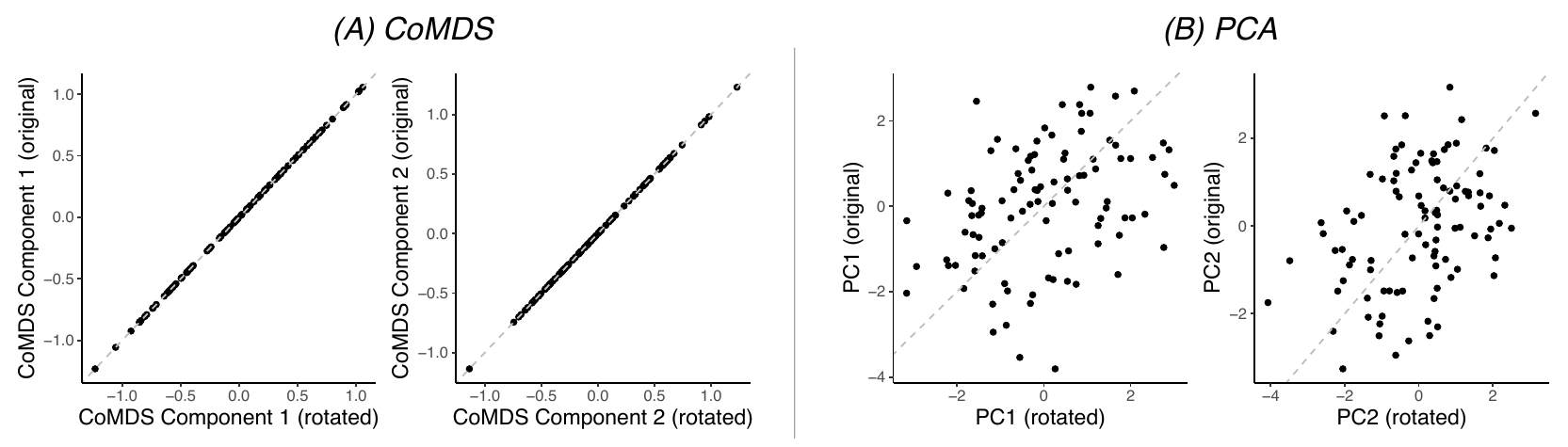}
    \caption{\textit{Rotation invariance simulation results.} (A) Scatter plots, comparing the estimated CoMDS consensus ideal points in the original (unrotated) scenario (y-axis) versus the rotated scenario (x-axis). The perfect correlation demonstrates CoMDS's invariance to rotations of the source ideal points. (B) Scatter plots, comparing the first two principal components (PCs) from PCA in the original (unrotated) scenario (y-axis) versus the rotated scenario (x-axis). The low correlation reveals PCA's sensitivity to rotations of the source ideal points.}
    \label{fig:toy_rotation_sim}
\end{figure}

In Figure~\ref{fig:toy_rotation_sim}A, we show that the first and second dimensions of the estimated CoMDS consensus ideal points $\bZh^{*}$ remain unchanged between the original (unrotated) scenario and the rotated scenario (as indicated by the perfect correlation). This is expected since the CoMDS optimization problem \eqref{eqn:optim} operates only on the pairwise distances between ideal points, and pairwise distances are invariant to shifts and rotations of the raw ideal point values. If additional rescalings were performed (although not applicable in this simulation), the source-specific weights $\bWh^{(1)}, \ldots, \bWh^{(S)}$ in \eqref{eqn:optim} would ensure that CoMDS is invariant to such rescalings of the raw ideal point values.

In contrast, Figure~\ref{fig:toy_rotation_sim}B shows that the first two principal components (PCs) from PCA change heavily depending on whether the source ideal points were rotated or not. This is indicated by the low correlation between PC1 (or PC2) scores from the original (unrotated) scenario and the rotated scenario. Here, the lack of invariance leads to a conundrum from an interpretation standpoint --- though the input source data in the two hypothetical scenarios are substantively equivalent, the resulting consensus PCs from the two hypothetical scenarios are different, thus raising questions about which consensus PCs are ``better'' and should be used for further analysis or interpretation.

\subsection{Imbalanced and correlated sources simulation}\label{app:simulations_imbalance}

By construction, CoMDS is specifically designed to identify patterns that are \textit{shared} across all source ideal point estimates. This design is well-aligned with the goal of estimating a consensus ideal point that reconciles multiple sources. General-purpose dimension reduction methods such as PCA, on the other hand, are not necessarily designed to find such consensus structures. In fact, PCA is designed to find the patterns that maximize the amount of variation in the data. These variance-maximizing patterns are often idiosyncratic (i.e., specific to a particular source) and not present in all input data sources. 

In practice, two common scenarios where the variance-maximizing patterns are likely to be idiosyncratic and found only in a single data source are (i) when one data source has a large number of features (or components) or (ii) when one data source has a high amount of correlation between its features (or components). Before illustrating this via simulations, we emphasize that such scenarios are practically relevant in the context of ideal point estimation. As an example of the imbalanced dimensions scenario, researchers may want to find the consensus between traditional ideal point estimates (e.g., NOMINATE and CF scores), which are 1- or 2-dimensional, and newer embeddings such as those generated from LLMs, which can easily output latent embeddings with hundreds of dimensions. Additionally, researchers may want to incorporate multiple variants of an ideal point estimation method (e.g., including different variants of CF scores \citep{bonica2014mapping, bonica2018rollcall, bonica_database_2024}), which are known to be highly correlated with each other.

\begin{figure}[tb]
    \centering
    \includegraphics[width=0.9\linewidth]{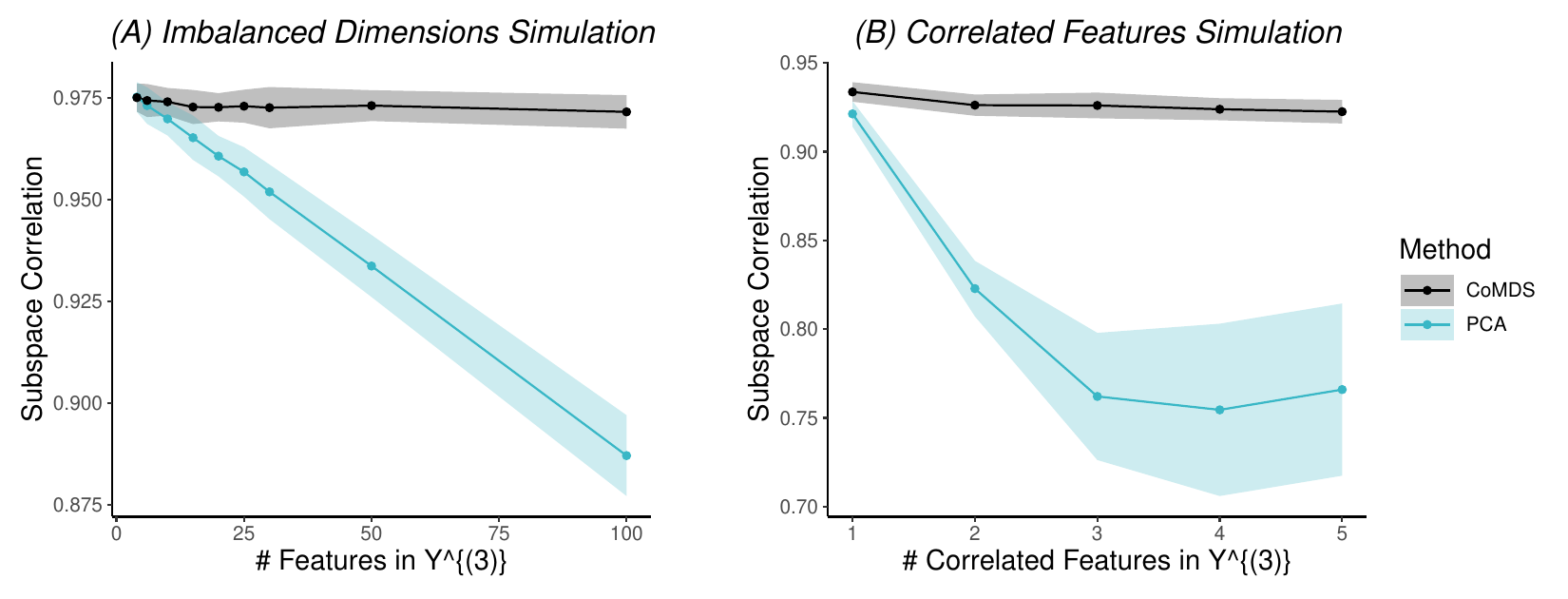}
    \caption{Subspace correlation metric between the true consensus ideal points $\bZ^{*}$ and the estimated consensus ideal points $\bZh^{*}$ (from CoMDS or PCA) in (A) the imbalanced dimensions and (B) the correlated features simulations. Higher correlations indicate better performance. While CoMDS accurately estimates the consensus ideal points under these simulated settings, PCA's performance declines as the dimensionality imbalance grows or as the amount of correlation within an idiosyncratic data source increases. Lines indicate the measured subspace correlation, averaged across 100 simulation replicates, and shaded areas represent $\pm 1SE$.}
    \label{fig:pca}
\end{figure}

\paragraph{Imbalanced dimensions simulation.} 
To illustrate the pitfalls of PCA under the imbalanced dimensions setting, we simulate two low-dimensional data sources $Y^{(1)}, Y^{(2)} \in \R^{300 \times 2}$ and one high-dimensional data source $Y^{(3)} \in \R^{300 \times p}$ as follows:
\begin{align*}
    Y^{(1)} &= \bZ^{*} \bD^{(1)} + \varepsilon^{(1)}\\
    Y^{(2)} &= \bZ^{*} \bD^{(2)} + \varepsilon^{(2)}\\
    Y^{(3)} &= [\bZ^{*}, \bZ^{(3)}] \bD^{(3)} + \varepsilon^{(3)},
\end{align*}
where $\bZ^{*} \in \R^{300 \times 2}$ is a standard normal matrix, encoding the true consensus structure; $\bZ^{(3)} \in \R^{300 \times (p - 2)}$ is a standard normal matrix, capturing the idiosyncrasies in the third data source; $\varepsilon^{(s)}_{ij} \stackrel{iid}{\sim} N(0, \sigma^2)$ is the additive noise term, where $\sigma$ is chosen to be half of the standard deviation of $\bZ^{*}$; and $\bD^{(1)}$, $\bD^{(2)}$, and $\bD^{(3)}$ are diagonal matrices that control the strengths of the consensus and idiosyncratic components in each data source. Specifically, we set $\bD^{(1)} = \bD^{(2)} = \text{diag\{4, 2\}} \in \R^{2 \times 2}$ and $\bD^{(3)} = \text{diag\{1.5, 1.5, 4, 2, 1, \ldots, 1\}} \in \R^{p \times p}$ so that the idiosyncratic signal dominates the consensus signal in the third (high-dimensional) data source. To assess the impact of the dimensionality imbalance, we vary the number of dimensions $p$ in the third data source from $p = 4$ to $100$ and repeat each simulation across 100 replicates.

In Figure~\ref{fig:pca}, we plot the correlation between the subspace induced by the true consensus ideal points $\bZ^{*}$ and the estimated consensus ideal points $\bZh^{*}$ (from CoMDS or PCA), where higher correlations indicate better performance. More specifically, we define the subspace correlation as the average squared singular value of the cross-product matrix between the true and estimated consensus subspaces \citep{bjorck1973numerical}. Intuitively, if the true and estimated consensus subspaces are well-aligned, then the subspace correlation will be close to 1. As seen in Figure~\ref{fig:pca}A, the performance of PCA deteriorates as the number of dimensions $p$ increases while CoMDS avoids this issue and yields an accurate estimate of the consensus ideal points $\bZ^{*}$ even as the dimension imbalance grows.

\paragraph{Correlated features simulation.} 
To illustrate the pitfalls of PCA under the correlated features setting, we simulate two data sources $Y^{(1)}, Y^{(2)} \in \R^{300 \times p}$ with independent features and one data source $Y^{(3)} \in \R^{300 \times p}$ with correlated features as follows:
\begin{align*}
    Y^{(1)} &= [\bZ^{*}, \bZ^{(1)}] \bD^{(1)} + \varepsilon^{(1)}\\
    Y^{(2)} &= [\bZ^{*}, \bZ^{(2)}] \bD^{(2)} + \varepsilon^{(2)}\\
    Y^{(3)} &= (\sqrt{1 - \omega}\; \bZ^{*} + \sqrt{\omega}\; \bZ^{(3)}) \mathbf{1}^{\top}_p \bD^{(3)} + \varepsilon^{(3)},
\end{align*} 
where $\bZ^{*} \in \R^{300 \times 1}$ is a standard normal matrix, encoding the true consensus structure; $\bZ^{(1)}, \bZ^{(2)} \in \R^{300 \times (p - 1)}$ and $\bZ^{(3)} \in \R^{300 \times 1}$ are standard normal matrices, capturing the idiosyncratic components in each data source; $\bD^{(1)} = \bD^{(2)} = \text{diag}\{1, \frac{0.2}{p-1}, \ldots, \frac{0.2}{p-1}\} \in \R^{p \times p}$ are diagonal matrices that control the strengths of the consensus and idiosyncratic components in the first and second data sources; $\bD^{(3)} = \text{diag}\{\frac{1.2}{p}, \ldots, \frac{1.2}{p}\} \in \R^{p \times p}$ so that the signal-to-noise ratio in $Y^{(3)}$ is similar to that in $Y^{(1)}$ and $Y^{(2)}$; $\omega$ controls the weight of the idiosyncrasies in the third data source; and $\varepsilon^{(s)}_{ij} \stackrel{iid}{\sim} N(0, \sigma^2)$ is the additive noise term, where $\sigma$ is chosen to be half of the standard deviation of $\bZ^{*}$. To assess the impact of the correlated features, we set $\omega = 0.9$, vary the amount of correlation in the third data source by varying $p$ from $p = 1, \ldots, 5$, and repeat each simulation across 100 replicates.

As shown in Figure~\ref{fig:pca}B, PCA's performance for estimating the consensus ideal points $\bZ^{*}$ deteriorates as the amount of correlation in the third data source increases. This issue occurs because PCA is designed to find the variance-maximizing patterns in the data, which in this case are dominated by the correlated features that are unique or idiosyncratic in the third data source. In contrast, CoMDS is able to effectively learn the underlying consensus ideal points $\bZ^{*}$ even as the amount of correlation increases in the third data source.

\subsection{Missing Data}\label{app:simulations_missing}

A final and important practical benefit of CoMDS is that it can still be applied even when some source ideal points are missing. Moreover, CoMDS can be applied without having to choose, use, and perform a missing data imputation method. Rather, in cases with missing values, terms in the CoMDS optimization problem \eqref{eqn:optim} that involve the missing source ideal points are given a weight of zero (via the missingness weights $\alpha_i^{(s)}$) and are thus effectively ignored in the estimation process. To better understand how the CoMDS estimates change in the presence of missing data, we provide an illustrative example, where we take the NOMINATE, CF, and campaign platform ideal point data used in Section~\ref{sec:application}, restricted to only those observations with data from all three sources, and we compare the estimated CoMDS consensus ideal points using the complete (non-missing) data versus using the data with a percentage (i.e., $1, 5, 10, 20\%$) of ideal points missing at random. 

\begin{figure}[tb]
    \centering
    \includegraphics[width=0.85\linewidth]{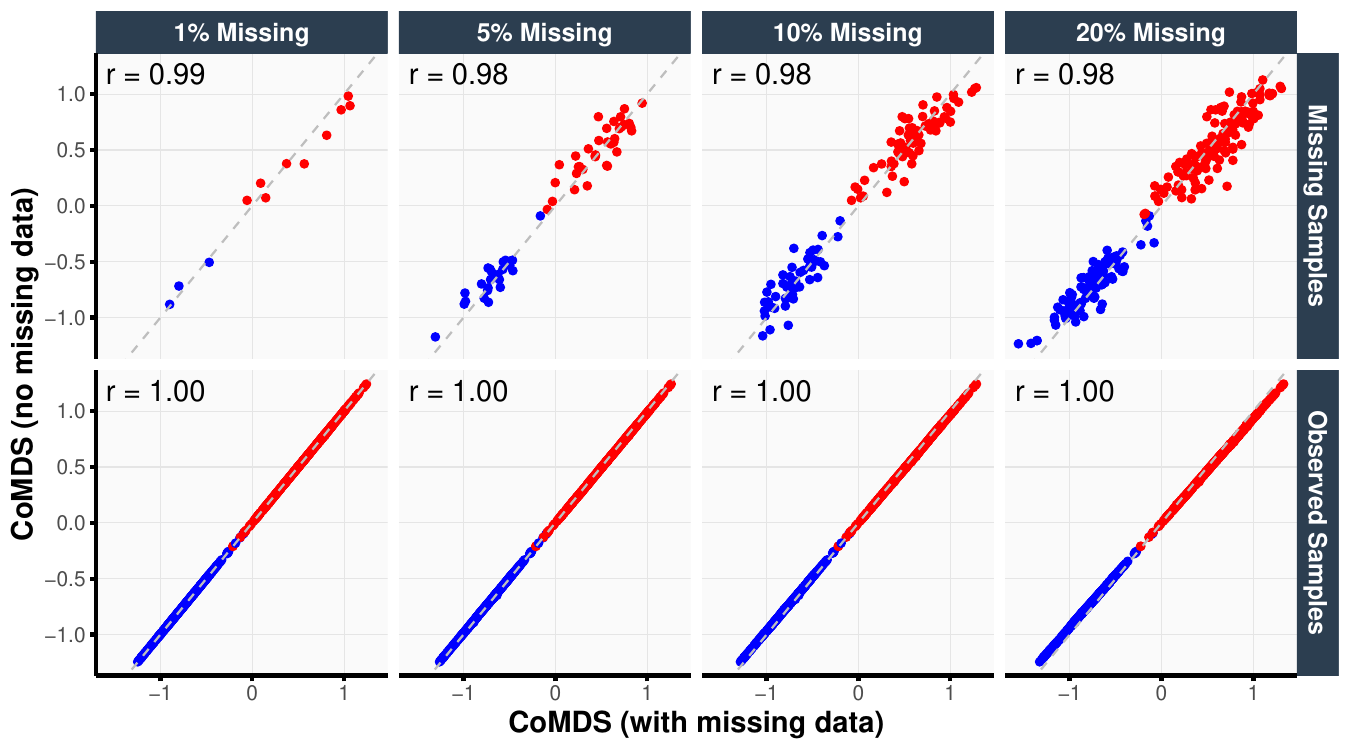}
    \caption{Comparison of the consensus ideal point estimates from CoMDS using the complete (non-missing) data (y-axis) versus using the data with a percentage of ideal points missing at random (x-axis). Samples with missing values are shown in the top row while samples with completely observed data are shown in the bottom row, and each column shows the results under a different percentage of missingness. The high correlations reveal that CoMDS provides similar estimates of the consensus ideal points regardless of whether we observed the complete data or whether some source ideal points were randomly missing.}
    \label{fig:missing}
\end{figure}

Figure~\ref{fig:missing} reveals that the estimated CoMDS consensus ideal points using the complete data (y-axis) are highly correlated with the estimated CoMDS consensus ideal points had a random proportion of the source ideal points been missing (x-axis). This correlation is essentially perfect when examining only the samples with complete data and $>0.98$ when examining only the samples with missing values. The strong concordance indicates that CoMDS consensus ideal points are similar (or stable) regardless of whether we observed the complete data or whether some source ideal points were randomly missing.

\begin{figure}[tb]
    \centering
    \includegraphics[width=0.85\linewidth]{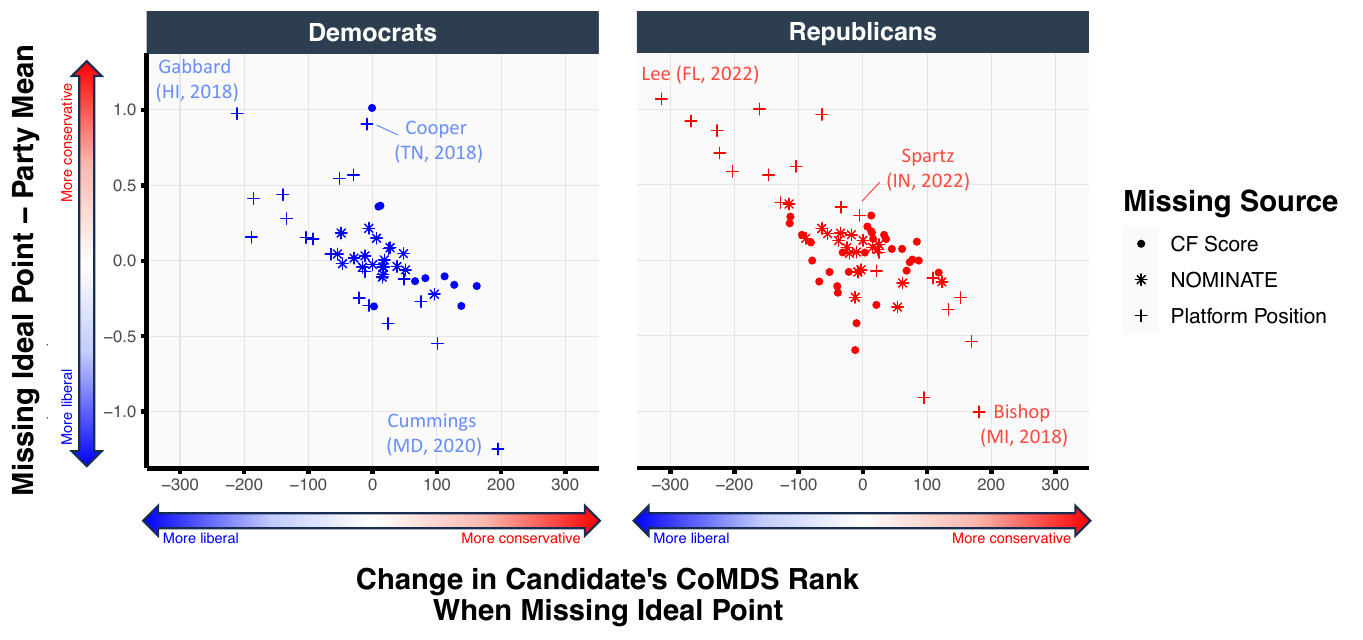}
    \caption{Stability across missing data imputation}
    \label{fig:missing_changes}
\end{figure}

\begin{table}[t]
\footnotesize
\begin{tabular}{lcccl}
\toprule
\textbf{Candidate} & \makecell{\textbf{NOMINATE} \\ \textbf{percentile}} & \makecell{\textbf{CF Score} \\ \textbf{percentile}} & \makecell{\textbf{Platform} \\ \textbf{percentile}} & \makecell[l]{\textbf{When \textit{missing} platform position,} \\ \textbf{the candidate's ideal point looks...}}\\
\midrule
\textcolor{blue}{Gabbard (D, HI, 2018)} & 79 & 63 & 98 & \textcolor{blue}{More liberal}\\ 
\textcolor{blue}{Cooper (D, TN, 2018)} & 97 & 95 & 97 & No change\\
\textcolor{blue}{Cummings (D, MD, 2020)} & 28 & 65 & 1 & \textcolor{red}{More conservative}\\
\textcolor{red}{Lee (R, FL, 2022)} & 25 & 32 & 97 & \textcolor{blue}{More liberal}\\
\textcolor{red}{Spartz (R, IN, 2022)} & 44 & 89 & 67 & No change\\
\textcolor{red}{Bishop (R, MI, 2018)} & 37 & 31 & 8 & \textcolor{red}{More conservative}\\
\bottomrule
\end{tabular}
\caption{}
\label{tab:missing_examples}
\end{table}

Still, there is some change, albeit generally small, in the estimated consensus ideal points when samples have missing source values. To gain intuition on how the missing data changes the estimated consensus ideal points, we more closely examined the candidates with missing values from the $10\%$ missingness case in Figure~\ref{fig:missing_changes} and Table~\ref{tab:missing_examples}. Beginning with Figure~\ref{fig:missing_changes}, we show on the x-axis the change in ranking of the candidate's consensus ideal point estimate when there was missingness (defined as the rank of the candidate's consensus ideal point estimate when using the missing data minus the rank of the candidate's consensus ideal point estimate when using the complete data, where rank $1$ indicates the most liberal candidate and rank $N$ indicates the most conservative candidate). This x-axis essentially captures whether the candidate's consensus ideal point became more liberal (negative values) or more conservative (positive values) when we were missing one of that candidate's ideal point sources. On the y-axis, we show the difference between the candidate's missing ideal point value and the candidate's party's mean ideal point value. This y-axis captures whether the candidate's missing ideal point is more liberal (negative values) or more conservative (positive values) than the typical candidate in their party. The key takeaway from Figure~\ref{fig:missing_changes} is that generally, when the missing ideal point value appears more conservative (or liberal) than average, then leaving out this ideal point will result in the candidate's consensus ideal point estimate to be more liberal (or conservative). Conversely, when the missing ideal point value appears more conservative (or liberal) than average, then generally including this ideal point in the CoMDS estimation will pull the candidate's consensus ideal point estimate towards being more conservative (or liberal), as one would expect. 

Furthermore, in Table~\ref{tab:missing_examples}, we highlight a few representative candidates from Figure~\ref{fig:missing_changes} and list the percentile ranks of their NOMINATE, CF, and platform ideal points. Larger percentile ranks indicate a more conservative-leaning ideal point while smaller percentile ranks indicate a more liberal-leaning ideal point. Each listed candidate was missing their platform ideal point, and as one might intuitively expect:
\begin{itemize}
    \item The ideal points for candidates, whose platform ideal points appeared more conservative than their observed NOMINATE and CF scores (e.g., Gabbard and Lee), appeared more liberal when their platform ideal point was missing. 
    \item The ideal points for candidates, whose platform ideal points were similar or central about their observed NOMINATE and CF scores (e.g., Cooper and Spartz), remained similar when their platform ideal point was missing.
    \item The ideal points for candidates, whose platform ideal points appeared more liberal than their observed NOMINATE and CF scores (e.g., Cummings and Bishop), appeared more conservative when their platform ideal point was missing.
\end{itemize}

\section{Additional discussion of CoMDS diagnostics}\label{app:relative_errors}

\setcounter{figure}{0}
\setcounter{table}{0}
\renewcommand{\thefigure}{C\arabic{figure}}
\renewcommand{\thetable}{C\arabic{table}}

As introduced in Section~\ref{subsec:diagnostics}, for a given source $s$, the relative error diagnostic measures the squared error between the pairwise distances in the $s^{th}$ source ideal point space and the pairwise distances in the estimated consensus ideal point space, relative to the total squared error across all $S$ sources. Generally speaking, this diagnostic measures how similar the source ideal point estimates are to the consensus ideal point estimates, with a larger relative error indicating greater dissimilarity. As a concrete example, a $20\%$ relative error indicates that $20\%$ of the total squared error loss across all sources is due to the $s^{th}$ source.

To provide additional intuition on how to interpret the relative error diagnostic measure, we conduct a brief simulation study, illustrating how the relative error diagnostic behaves as we vary the strength of the idiosyncratic signal in the source ideal points. Specifically, we simulate three data sources $Y^{(1)}, Y^{(2)}, Y^{(3)} \in \R^{300 \times 2}$ via:
\begin{align*}
    Y^{(1)} &= \bZ^{*} + \varepsilon^{(1)}\\
    Y^{(2)} &= \bZ^{*} + \varepsilon^{(2)}\\
    Y^{(3)} &= \sqrt{1 - \omega}\;\bZ^{*} + \sqrt{\omega} \bZ^{(3)} + \varepsilon^{(3)},
\end{align*}
where $\bZ^{*} \in \R^{300 \times 2}$ is a standard normal matrix that has been orthogonalized (via the Gram-Schmidt procedure) and encodes the true consensus structure; $\bZ^{(3)} \in \R^{300 \times 2}$ is a standard normal matrix that has been orthogonalized and encodes the idiosyncratic component in the third data source; $\varepsilon^{(s)}_{ij} \stackrel{iid}{\sim} N(0, \sigma^2)$ is the additive noise term, where $\sigma$ is chosen to be half of the standard deviation of $\bZ^{*}$; and $\omega$ controls the strength of the idiosyncratic signal in the third data source. At a high-level, the first two data sources $Y^{(1)}$ and $Y^{(2)}$ contain only the consensus information with no idiosyncratic component while the third data source $Y^{(3)}$ is a weighted combination of both the consensus and idiosyncratic components.

\begin{figure}[tb]
    \centering
    \includegraphics[width=0.55\linewidth]{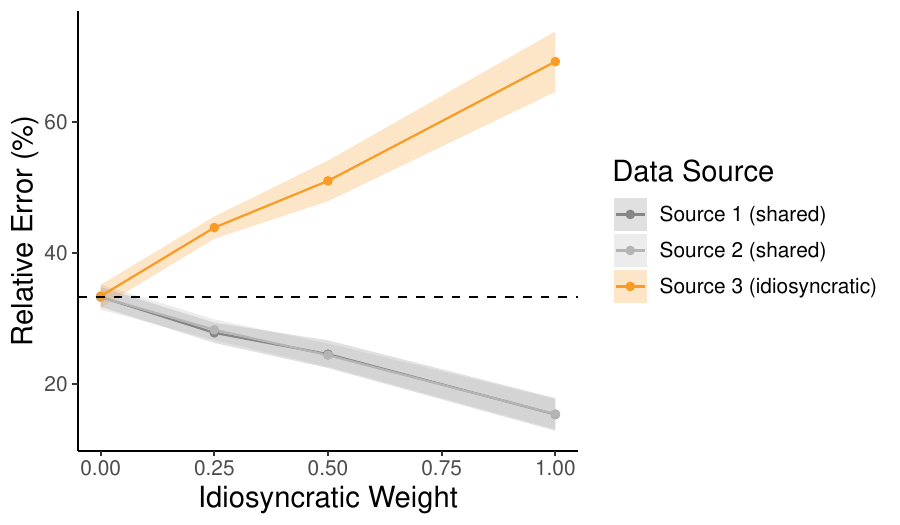}
    \caption{Relative errors of each source as we vary the idiosyncratic strength $\omega$ in the third source $Y^{(3)}$. When there are no idiosyncrasies, each source contributes equally to the total squared error loss and exhibits equal relative errors. As the idiosyncratic strength increases, the relative error of the third source increases while the relative errors of the first two sources decrease.}
    \label{fig:relative_errors_sim}
\end{figure}

In Figure~\ref{fig:relative_errors_sim}, we summarize each source's relative error as we vary $\omega$, or the idiosyncratic strength in $Y^{(3)}$. Unsurprisingly, when there is no idiosyncratic signal in $Y^{(3)}$ (i.e., $\omega = 0$), all three data sources contain only the consensus information, and thus, the relative error is equal across all three sources. In other words, each source contributes equally (i.e., $1/3 = 33.3\%$) to the total squared error loss, and the consensus ideal point estimates are a perfect consensus or equally similar to each source's ideal point estimates. 

As the idiosyncratic signal in $Y^{(3)}$, controlled by $\omega$, increases, the relative error of the third source $Y^{(3)}$ increases while the relative errors of the first two sources $Y^{(1)}$ and $Y^{(2)}$ decrease. This is to say that the estimated CoMDS consensus ideal points grow increasingly more different from the third source $Y^{(3)}$ as $\omega$ increases. This is intuitive since a larger $\omega$ corresponds to a stronger idiosyncratic signal in $Y^{(3)}$, thereby making it less similar to the true underlying consensus ideal points $\bZ^{*}$.

\section{Diagnostics and stability analysis of candidate positioning application}
\label{sec:pcs}

\setcounter{figure}{0}
\setcounter{table}{0}
\renewcommand{\thefigure}{D\arabic{figure}}
\renewcommand{\thetable}{D\arabic{table}}

To supplement our candidate position study from Section~\ref{sec:application}, we report the recommended CoMDS diagnostics (see Section~\ref{subsec:diagnostics}) and conduct an extensive stability analysis \citep{yu2020veridical} to ensure that the presented findings are stable and robust to alternative, but equally-reasonable analysis choices. Here, key choices in the candidate positioning analysis include (i) which primary sources to include in the CoMDS estimation, (ii) whether to include only the first component of NOMINATE or both of its components, (iii) whether to use the classic CF scores or more sophisticated variants of CF scores, and (iv) whether or not to exclude candidates with missing source ideal point estimates. In what follows, we evaluate how each of these choices impact the estimated consensus ideal points from CoMDS and ultimately demonstrate that the estimated consensus ideal points shown in Section~\ref{sec:application} are stable and robust to these different modeling choices. Importantly, this demonstrated stability helps to enhance the reliability and validity of our substantive findings.

\begin{figure}[tb]
    \centering
    \includegraphics[width=0.95\linewidth]{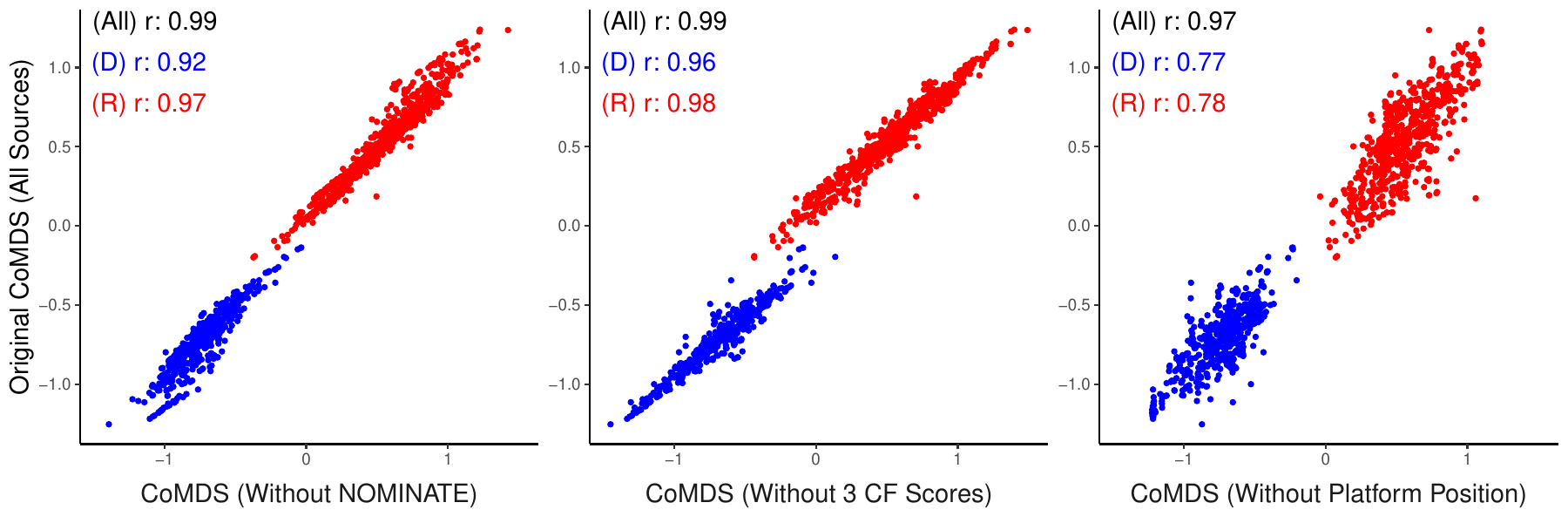}
    \caption{Leave-one-out stability analysis. We compare the original CoMDS consensus ideal points (y-axis) with the re-estimated CoMDS consensus ideal points when leaving out each source type (x-axis). The high correlations indicate that the estimated consensus ideal points are stable and robust to the exclusion of any single source type.}
    \label{fig:loo}
\end{figure}

\paragraph{CoMDS diagnostics.} As recommended in Section~\ref{subsec:diagnostics}, we perform a leave-one-out analysis and assess the stability of the estimated consensus ideal points (fitted on the candidates with no missingness) when leaving out each data source one at a time. Figure~\ref{fig:loo} compares the CoMDS consensus ideal points (y-axis), which included three types of input sources (i.e., NOMINATE, the three CF scores, and platform positions), with the re-estimated CoMDS consensus ideal points (x-axis) when leaving out each source type. We observe that the correlation between the original and re-estimated consensus ideal points is slightly lower (albeit still very high) when leaving out the platform positions compared to leaving out NOMINATE or the three CF scores. This suggests that the platform positions provide some distinctive information that may be overlooked when only considering NOMINATE and CF scores. 

\begin{figure}[H]
    \centering
    \includegraphics[width=0.5\linewidth]{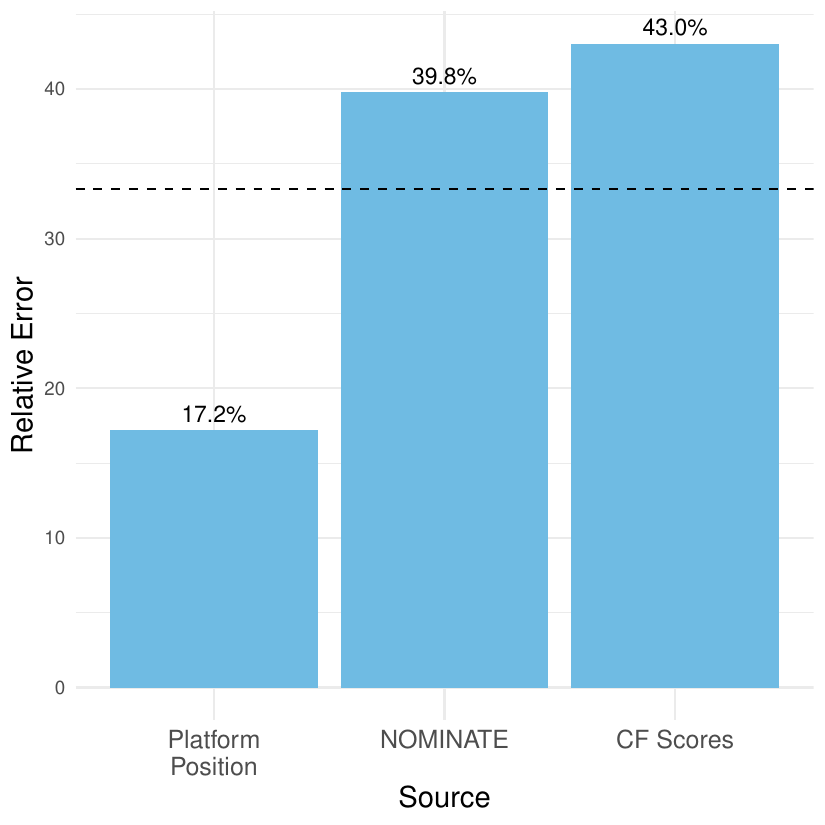}
    \caption{Relative error diagnostic}
    \label{fig:relative-error}
\end{figure}

At the same time, the relative error diagnostic in Figure~\ref{fig:relative-error} reveals that when analyzing the platform positions, NOMINATE, and CF scores simultaneously via CoMDS, the platform positions have the lowest relative error among the three source ideal point types. This indicates that the shared information across all source measures is more similar to the platform positions than to NOMINATE or CF scores, or conversely, that the amount of remaining idiosyncratic variation, which is orthogonal to the shared component, is lower among platform positions relative to CF scores and NOMINATE. Taken together, these diagnostics suggest that while the platform positions provide distinctive information that is not fully captured by NOMINATE or CF scores, they are closely aligned with the shared component captured by the consensus ideal points.

\begin{figure}[tb]
    \centering
    \includegraphics[width=1\linewidth]{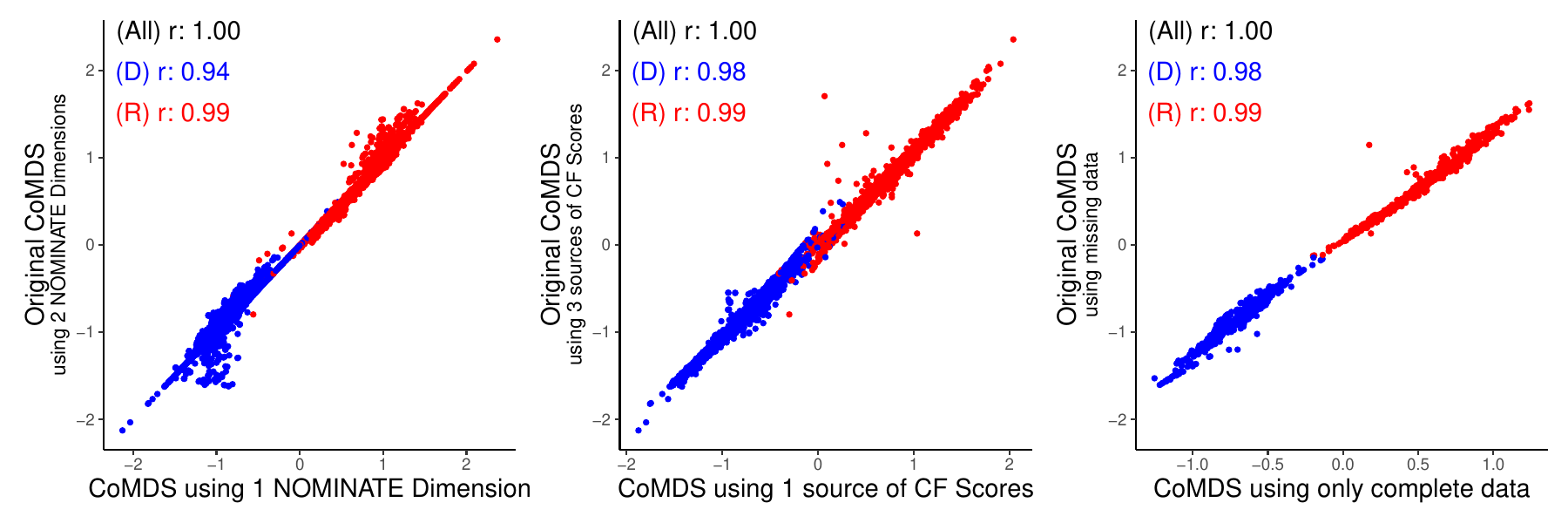}
    \caption{Stability across modeling choices. We compare the original CoMDS consensus ideal points (y-axis) with the re-estimated CoMDS consensus ideal points when using only the first NOMINATE component (left), one source of CF scores (middle), and only including candidates with complete data across all sources (right). The high correlations indicate that the estimated consensus ideal points are stable and robust to these different modeling choices.}
    \label{fig:stability}
\end{figure}

\paragraph{Stability across other modeling choices.} Beyond the choice of input sources, we further assess the stability of the CoMDS consensus ideal point estimates across other modeling choices. Specifically, while we chose to use both NOMINATE components and three variants of CF scores\footnote{Each CF score source is given a $\frac{1}{3}$ weight so as to not overly-emphasize CF scores relative to NOMINATE and the platform positions.} in the original CoMDS analysis (namely, the classic CF scores, the dynamic variant, and DW-DIME \citep{bonica2014mapping, bonica2018rollcall, bonica_database_2024}), we could have used only the first NOMINATE component or included only the classical variant of CF scores. The left two panels of Figure~\ref{fig:stability} reveal that these alternative modeling choices yield very similar consensus ideal point estimates, as indicated by the high correlations. The rightmost panel of Figure~\ref{fig:stability} compares the consensus ideal estimates when CoMDS was estimated using only the subset of candidates with complete data (i.e., no missing source ideal points) as opposed to using the larger set of candidates who have at least two of the three sources available (as used in the main analysis). Similar to before, this comparison revealed a high correlation, indicating that this choice does not substantially change the estimated consensus ideal points.

\begin{figure}[tb]
    \centering
    \includegraphics[width=0.95\linewidth]{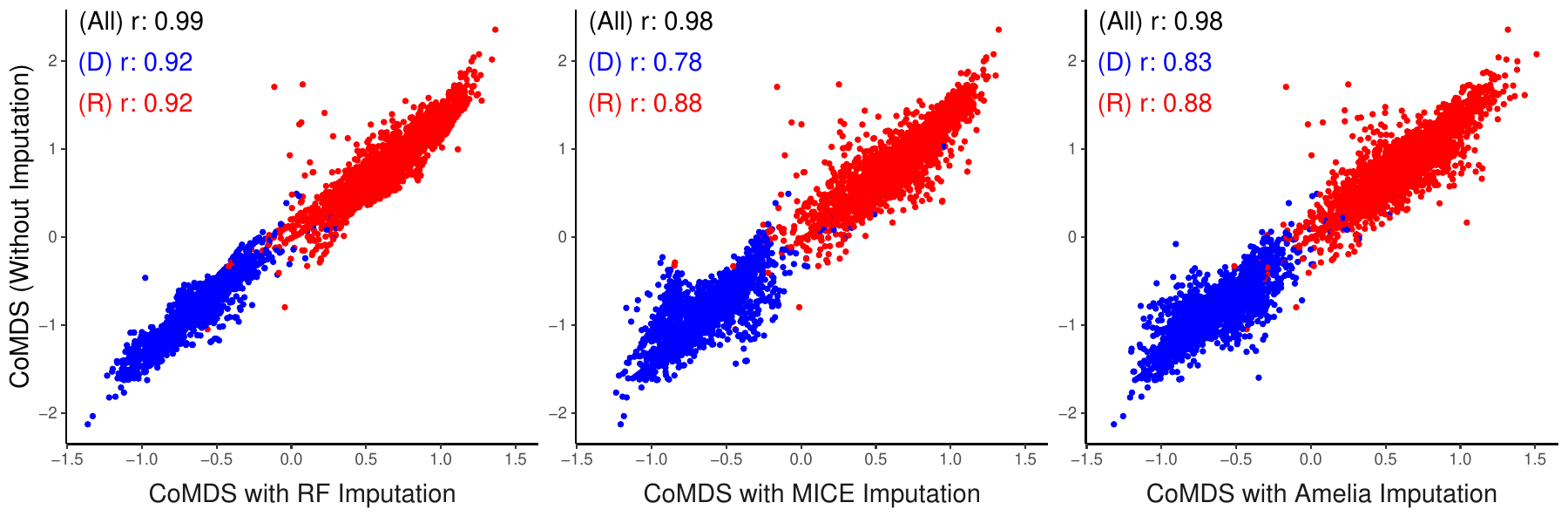}
    \caption{Stability between original CoMDS consensus ideal points without imputation (y-axis) and the CoMDS estimates after performing missing data imputation (x-axis) using RF-based imputation with `MissForest` (left), MICE imputation (middle), and Amelia imputation (right). The high correlations indicate that the estimated consensus ideal points are stable and robust with respect to the handling of missing data.}
    \label{fig:stability_imputation}
\end{figure}

\begin{figure}[tb]
    \centering
    \includegraphics[width=0.85\linewidth]{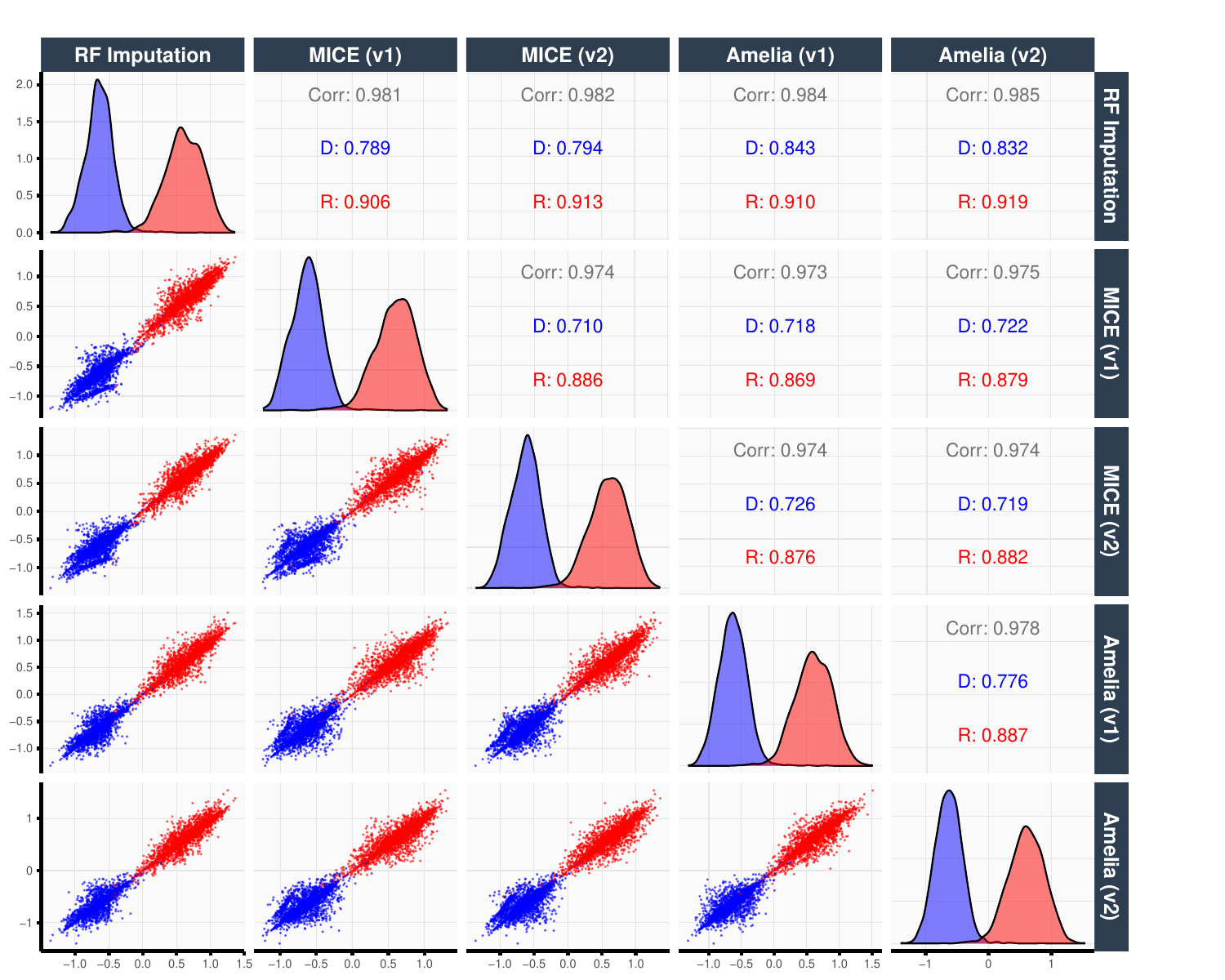}
    \caption{Stability of CoMDS ideal point estimates when using different missing data imputation methods: RF-based imputation with `MissForest`, two random draws from MICE imputation (labeled v1 and v2), and two random draws from Amelia imputation (labeled v1 and v2).}
    \label{fig:stability_imputation_pair}
\end{figure}

\paragraph{Stability across missing data imputation choices.}
Recall that in the analysis including candidates with possibly missing source ideal points, we applied CoMDS without imputing the missing values since CoMDS can naturally handle missing data by setting their missingness weights $\alpha_i^{(s)}$ to zero. In Figure~\ref{fig:stability_imputation}, we compare these CoMDS consensus ideal point estimates to what we would have obtained had we first imputed the missing source ideal points using various data imputation methods and then applied CoMDS to this imputed dataset. Since different imputation methods generally impose different assumptions on both the data-generating process and the missingness mechanism, we considered three imputation methods that are popularly used in practice: RF-based imputation using \texttt{missForest} \citep{stekhoven2012missforest}, multivariate imputation by chained equations (MICE) \citep{van2011mice}, and Amelia \citep{honaker2011amelia}. From Figure~\ref{fig:stability_imputation}, we see that consensus ideal point estimates from CoMDS (without imputation) are highly correlated with the CoMDS estimates obtained after imputing the missing values no matter the choice of imputation method, though the correlation with the RF-based imputation is highest. We also show the correlation between the CoMDS consensus ideal point estimates when using different data imputation methods or different runs of the same data imputation method in Figure~\ref{fig:stability_imputation_pair}. Interestingly, the correlations observed between CoMDS estimates using the same imputation method but a different random draw (Figure~\ref{fig:stability_imputation_pair}) are frequently lower than the correlations observed between CoMDS without imputation versus with different imputation methods (Figure~\ref{fig:stability_imputation}). This is to illustrate that data imputation methods often introduce more instabilities into an analysis, complicating the substantive interpretation. Further, it underscores the benefit of CoMDS's ability to naturally handle missing values without the need for an imputation step during data preprocessing.

\clearpage

\section{Additional candidate positioning analyses} \label{app:robustness}

\setcounter{figure}{0}
\setcounter{table}{0}
\renewcommand{\thefigure}{E\arabic{figure}}
\renewcommand{\thetable}{E\arabic{table}}

To supplement the substantive results reported in the paper, here we present additional analyses which may be of interest given other possible sample selections, aggregation approaches, or modeling choices. In particular, we compare our CoMDS estimates of House candidates' ideal points to estimates obtained with alternative aggregation methods, subset to complete cases (i.e. no missingness across source measures), and assess CoMDS estimates based on just the three main variants of source measures (i.e. NOMINATE first-dimension, classic CF scores, and campaign platform positions).

\subsection{Comparisons to alternative methods}

As discussed in Appendices \ref{app:ted} and \ref{app:simulations}, the aggregation approaches most closely related to CoMDS are MD2S and PCA, although crucial differences between all three are highly relevant in our particular application. To assess how different the methods are in practice, when applied to our substantive context, we re-scaled candidates' aggregate ideal points using the same sources of data with the only difference being the use of MD2S or PCA instead of CoMDS. Figure \ref{fig:corr-alt} plots relationships between CoMDS-based ideal points and the alternative ideal points. Unlike CoMDS, MD2S and PCA cannot handle missingness directly, so we use random forest imputation\footnote{RF imputation was chosen to maximize (potential) similarity with CoMDS as Figure \ref{fig:stability_imputation} shows that this was the imputation method which resulted in estimates most similar to those reached with our non-imputed CoMDS approach.} and Bayesian PCA, respectively, while also showing relationships among complete cases where differences in the handling of missingness should play less of a role in driving correlations downwards.

Figure \ref{fig:corr-alt} demonstrates strong but imperfect correlations between estimates produced by CoMDS versus alternative aggregation approaches. Even within party, most correlations remain well above 0.5. The results suggest that, at least within our application and sample, Bayesian PCA provides estimates which are more similar to CoMDS than MD2S. Imperfect correlations between estimates are likely due in part to alternative methods' implicit overweighting of correlated source measures, as well as their imputation approaches which affect the underlying space in which even the subset of complete cases were originally scaled. 

While these results showcase the relationships between estimates of House candidates' ideal points reached with different methods, holding constant the underlying sources of data, an existing, off-the-shelf composite measure which constitutes an attractive alternative for applied researchers is found in \citet{bonica_database_2024}. This approach applies Amelia imputation and PCA to a variety of roll-call and contribution-based measures, which means that such estimates differ from those obtained with CoMDS along both source data and methodological dimensions. Given composite scores' source data, it is unsurprising that Table E1 shows that using them in substantive analyses in the domains of roll-call voting and campaign contributions produce results which are highly similar to those with domain-specific NOMINATE and CF scores, respectively, reported in the main results. Conversely, there is little relationship between the composite score and lexical diversity, which is consistent with DW-NOMINATE and CF scores' lack of significant relationships with lexical diversity.

\begin{figure}[t!]
    \centering
    \caption{Relationships Between Ideal Points Estimated via CoMDS and Alternative Aggregation Methods}
    \includegraphics[width=\linewidth]{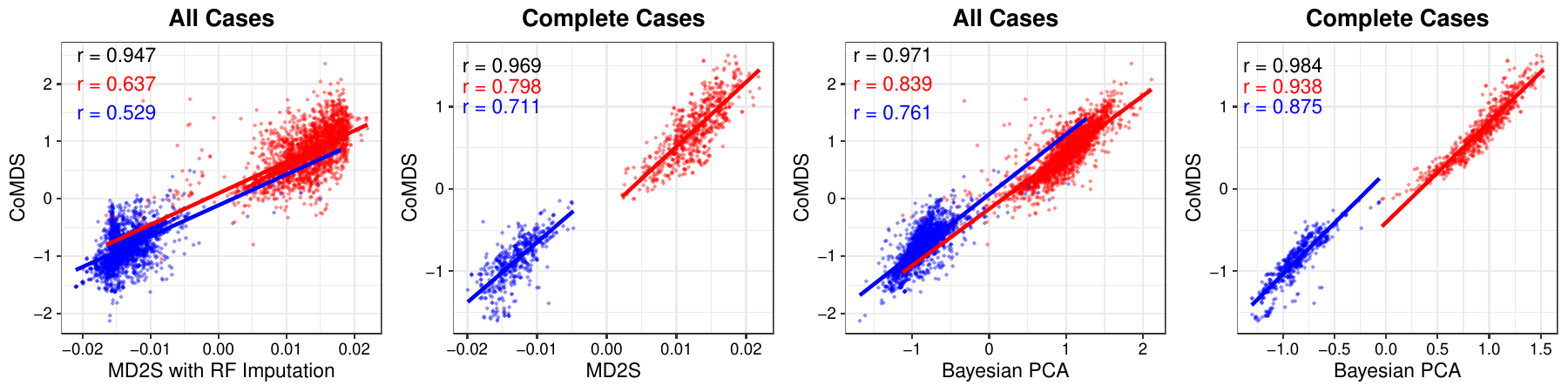}
\end{figure}\label{fig:corr-alt}

\begin{table}
\centering\centering
\caption{Main Results with Bonica Composite Score}
\centering
\fontsize{10}{12}\selectfont
\begin{tabular}[t]{lcccccc}
\toprule
\multicolumn{1}{c}{ } & \multicolumn{2}{c}{Roll-Call Disloyalty} & \multicolumn{2}{c}{\# Campaign Donors} & \multicolumn{2}{c}{Lexical Diversity} \\
\cmidrule(l{3pt}r{3pt}){2-3} \cmidrule(l{3pt}r{3pt}){4-5} \cmidrule(l{3pt}r{3pt}){6-7}
  & Dems & Reps & Dems  & Reps  & Dems   & Reps  \\
\midrule
Composite Score & \num{ 0.793}*** & \num{-0.737}*** & \num{ 0.682}*** & \num{-0.757}*** & \num{-0.191} & \num{ 0.103}\\
 & (\num{0.045}) & (\num{0.055}) & (\num{0.094}) & (\num{0.073}) & (\num{0.106}) & (\num{0.075})\\
\midrule
Observations & \num{1,208} & \num{1,345} & \num{2,628} & \num{2,759} & \num{1,835} & \num{1,993}\\
\bottomrule
\multicolumn{7}{l}{\rule{0pt}{1em}* p $<$ 0.05, ** p $<$ 0.01, *** p $<$ 0.001}\\
\end{tabular}
\end{table}

\subsection{Robustness to including only complete cases}

Given CoMDS' ability to handle missing data directly, the main analyses based on consensus ideal points include cases where a candidate had nonmissing estimates for two out of the three main source measures. However, as detailed in Appendix \ref{app:simulations}, missingness will affect estimates to the extent that a candidate's missing source estimate would differ from her nonmissing source estimates. Moreover, due to the nature of the source measures we include, the target universe of each differ systematically, such that each measure has highly nonrandom missingness. For example, NOMINATE relies upon congressional roll-call records and therefore excludes all candidates who fail to win election to Congress, a population that differs systematically from those who win election. Here, we re-run all analyses to assess consensus ideal points among the subset of complete cases, i.e. legislators with campaign website platforms.

\begin{figure}[t!]
    \centering
    \caption{Correlations Between Source Measures Among Complete Cases}
    \includegraphics[width=\linewidth]{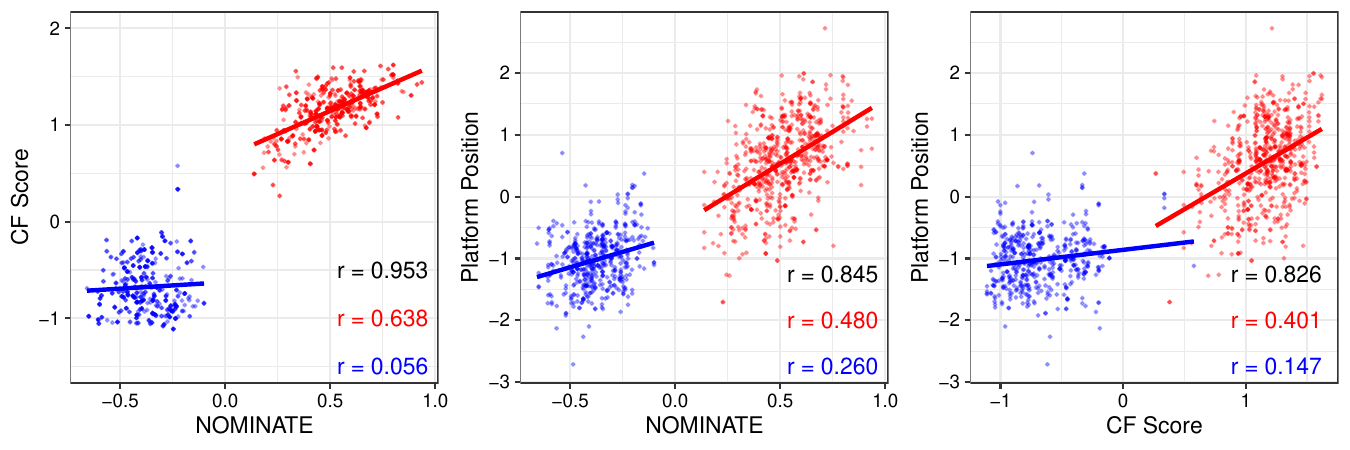}
    \label{fig:corr-complete}
\end{figure}

\begin{figure}[t!]
    \centering
    \caption{Distribution of Consensus Ideal Point Among Complete Cases}
    \includegraphics[width=.7\linewidth]{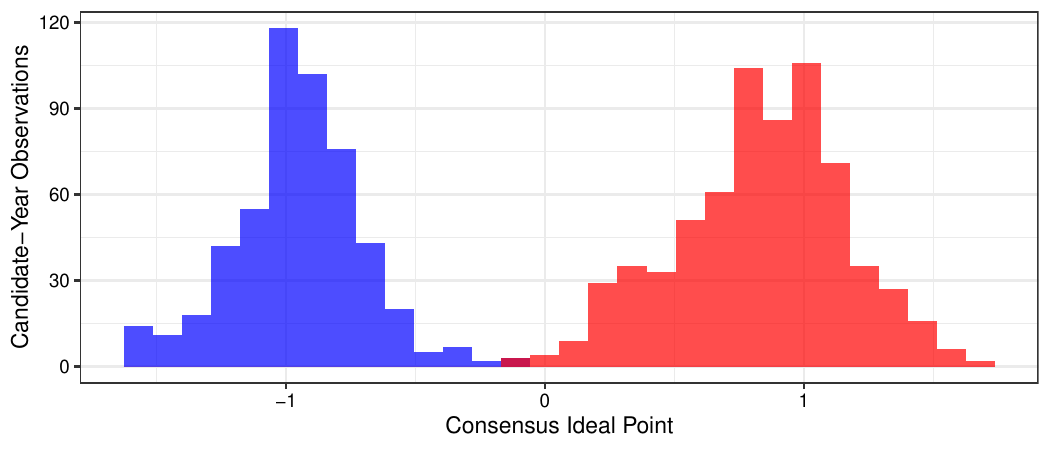}
        \label{fig:hist-complete}
\end{figure}

Figure \ref{fig:corr-complete} shows modestly higher correlations among the source measure estimates of these complete cases than among all of the cases included in the main analyses. The largest increases are evident for correlations between CF scores and platform positions, which is relatively unsurprising given that both measures cover a large number of nonincumbent candidates, all of whom are dropped from the complete case analysis by virtue of missing a NOMINATE score. Correlations between NOMINATE and CF scores of complete cases, i.e. those who have campaign platforms, increase only slightly, and correlations between NOMINATE and platform positions remain identical as all candidates with both likewise have a CF score. Additionally, Figure \ref{fig:hist-complete} suggests that the distribution of consensus ideal points among complete cases is relatively similar to that among all cases included in the main sample, albeit with slightly less overlap between parties.

\begin{figure}[t!]
    \centering
    \caption{Relationships Between Source and Consensus Measures Among Complete Cases}
    \includegraphics[width=\linewidth]{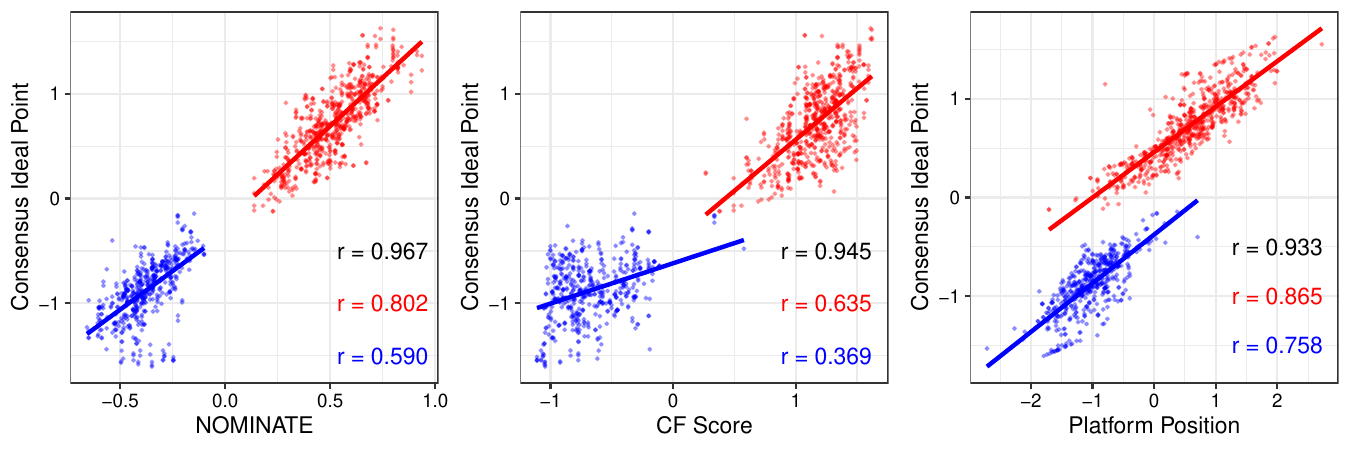}
    \label{fig:compare-complete}
\end{figure}

Additionally, Figure \ref{fig:compare-complete} shows that the correlations between consensus ideal points and CF scores become modestly stronger while correlations between consensus ideal points and NOMINATE and platform positions become modestly weaker when subsetting to complete cases. However, relationships with NOMINATE and platform positions nevertheless remain stronger than relationships with CF scores among complete cases. This suggests that the universe of complete cases have consensus ideal points that are more strongly related to their CF scores compared to the universe of cases with missingness, but consensus ideal points nevertheless exhibit stronger relationships with the other two measures among both complete and incomplete cases.

Table E2 re-estimates the main substantive results based on the consensus ideal points among complete cases only. The roll-call disloyalty estimates for both Democrats and Republicans are qualitatively similar yet smaller among complete cases, suggesting even greater overestimation of the domain-specific measure in this context. The campaign donor results suggest little meaningful relationship between consensus ideal point and financial base among complete cases. This is consistent with the highly mixed results across complete cases based on source measures reported in Table 2. 
Finally, the lexical diversity results based on consensus ideal points are even smaller among complete cases. Overall, subsetting the substantive results to complete cases suggests that domain-specific estimates are, if anything, even more overstated compared to consensus ideal points. 

\begin{table}
\centering\centering
\caption{Consensus Ideal Point Results Among Complete Cases}
\centering
\fontsize{10}{12}\selectfont
\begin{tabular}[t]{lcccccc}
\toprule
\multicolumn{1}{c}{ } & \multicolumn{2}{c}{Roll-Call Disloyalty} & \multicolumn{2}{c}{\# Campaign Donors} & \multicolumn{2}{c}{Lexical Diversity} \\
\cmidrule(l{3pt}r{3pt}){2-3} \cmidrule(l{3pt}r{3pt}){4-5} \cmidrule(l{3pt}r{3pt}){6-7}
  & Dems & Reps & Dems  & Reps  & Dems   & Reps  \\
\midrule
Consensus & \num{ 0.574}*** & \num{-0.437}*** & \num{0.002} & \num{-0.088} & \num{-0.023} & \num{-0.301}*\\
 & (\num{0.050}) & (\num{0.044}) & (\num{0.080}) & (\num{0.056}) & (\num{0.208}) & (\num{0.141})\\
\midrule
Observations & \num{493} & \num{660} & \num{516} & \num{678} & \num{516} & \num{678}\\
\bottomrule
\multicolumn{7}{l}{\rule{0pt}{1em}* p $<$ 0.05, ** p $<$ 0.01, *** p $<$ 0.001}\\
\end{tabular}
\end{table}

\FloatBarrier

\subsection{Robustness to scaling three main measures}

\begin{figure}[]
    \centering
    \caption{Distribution of Consensus Ideal Point Scaled with Three Main Measures}
    \includegraphics[width=.7\linewidth]{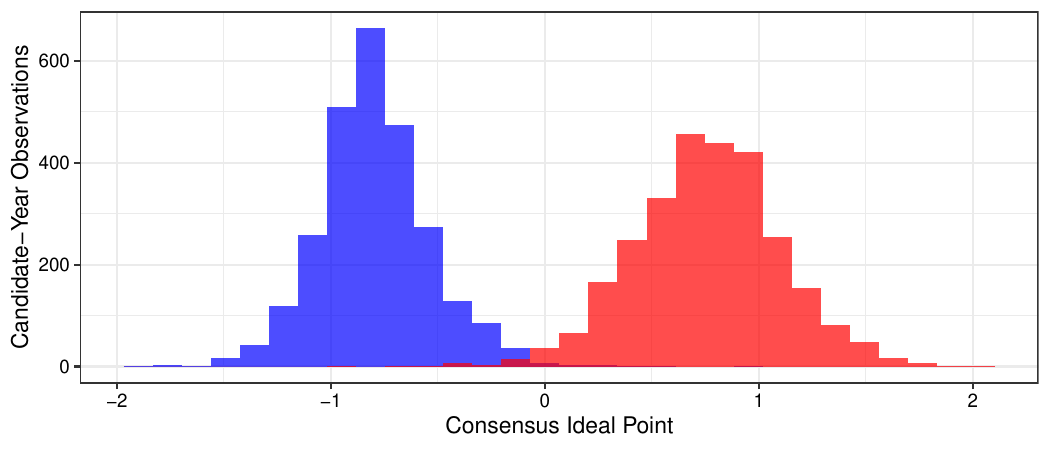}
    \label{fig:hist-measures3}
\end{figure}

\begin{figure}[]
    \centering
    \caption{Relationships Between Source and Consensus Measures Based on Three Main Measures}
    \includegraphics[width=\linewidth]{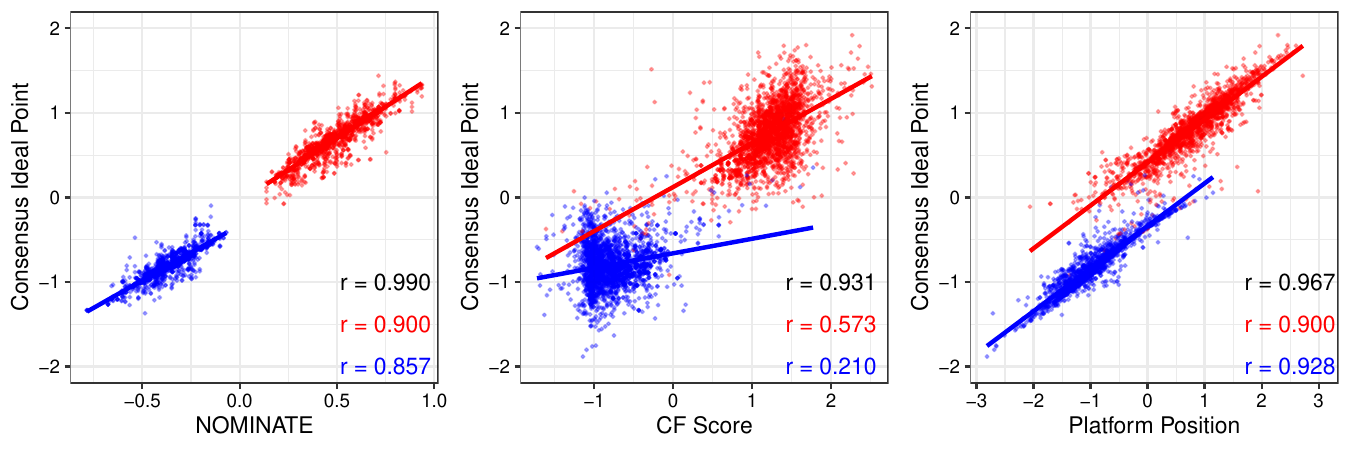}
        \label{fig:compare-measures3}
\end{figure}

When using CoMDS to scale the positions of House candidates, we relied upon both dimensions of NOMINATE and the classic, dynamic, and NOMINATE-targeted variants of CF scores, in addition to unidimensional platform positions. Scholars typically focus on the first dimension of NOMINATE as well as classic CF scores, however. Here, we re-estimate candidates' ideal points using just these three main measures, and replicate all substantive results with the estimates based on three such measures. Figure \ref{fig:hist-measures3} suggests that the distribution of consensus ideal points based on three main measures is relatively similar to the distribution based on the original set of measures, aside from somewhat less spread among Democrats' consensus ideal points. Likewise, Figure \ref{fig:compare-measures3} confirms that relationships between each of the source measures and the consensus ideal points look relatively similar regardless of whether the three main source measures versus the original set of source measures are used. Finally, Table E3 reports substantive results using consensus ideal points estimated with the three main source measures. Overall, takeaways remain qualitatively similar to those from the consensus ideal points which include other variants of source measures aside from the lexical diversity results among Democrats, which become essentially equivalent to the results originally reached using platform positions.

\begin{table}[t!]
\centering\centering
\caption{Relationship Between Consensus Ideal Points Based on Three Main Measures and Substantive Variables}
\centering
\fontsize{10}{12}\selectfont
\begin{tabular}[t]{lcccccc}
\toprule
\multicolumn{1}{c}{ } & \multicolumn{2}{c}{Partisan Disloyalty} & \multicolumn{2}{c}{\# Campaign Donors} & \multicolumn{2}{c}{Lexical Diversity} \\
\cmidrule(l{3pt}r{3pt}){2-3} \cmidrule(l{3pt}r{3pt}){4-5} \cmidrule(l{3pt}r{3pt}){6-7}
  & Dems & Reps & Dems  & Reps  & Dems   & Reps  \\
\midrule
Consensus & \num{ 0.631}*** & \num{-0.556}*** & \num{-0.260}*** & \num{-0.475}*** & \num{ 0.478}*** & \num{-0.538}***\\
 & (\num{0.034}) & (\num{0.032}) & (\num{0.059}) & (\num{0.057}) & (\num{0.080}) & (\num{0.063})\\
\midrule
Observations & \num{1,206} & \num{1,340} & \num{2,629} & \num{2,760} & \num{1,836} & \num{1,994}\\
\bottomrule
\multicolumn{7}{l}{\rule{0pt}{1em}* p $<$ 0.05, ** p $<$ 0.01, *** p $<$ 0.001}\\
\end{tabular}
\end{table}

\FloatBarrier

\subsection{Additional analyses} 

\begin{figure}[H]
    \centering
    \caption{Ideal Points and Roll-Call Partisan Disloyalty}
    \includegraphics[width=.7\linewidth]{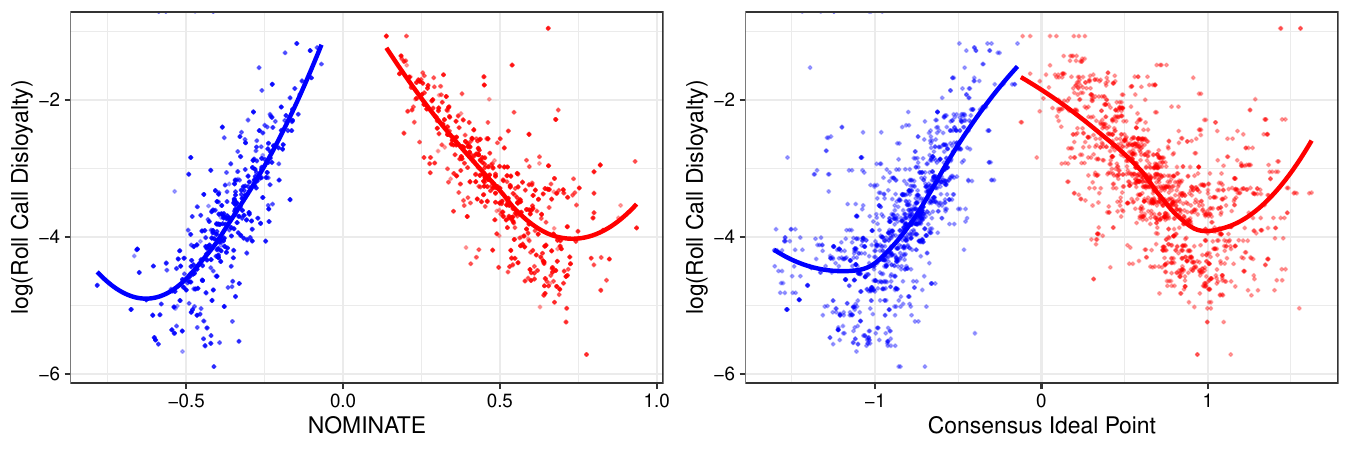}
    \label{fig:partydisloyalty}
\end{figure}

\begin{figure}[H]
    \centering
        \caption{Ideal Points  and Fundraising Success}
    \includegraphics[width=.7\linewidth]{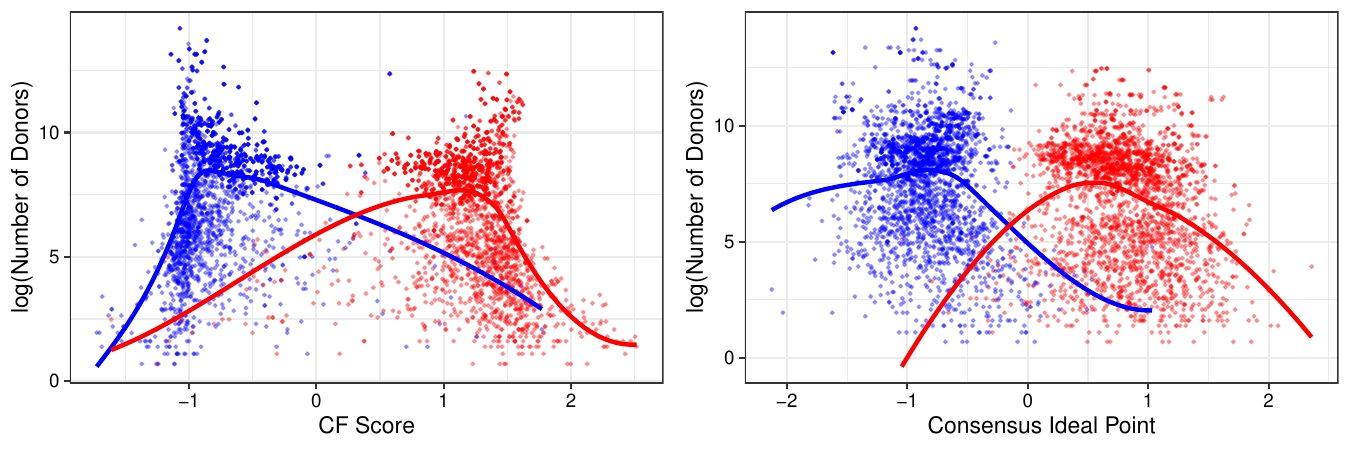}
    \label{fig:ndonors}
\end{figure}

\begin{figure}[H]
    \centering
        \caption{Ideal Points  and Lexical Diversity}
    \includegraphics[width=.7\linewidth]{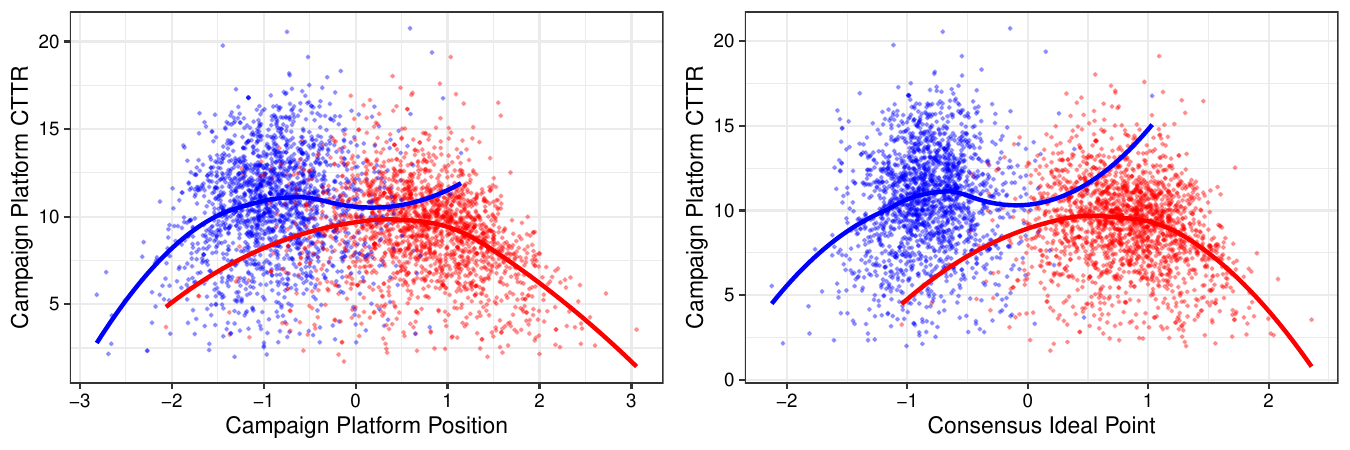}
    \label{fig:lexdiv}
\end{figure}

Figures \ref{fig:partydisloyalty}, \ref{fig:ndonors}, and \ref{fig:lexdiv} present scatterplots of the basic bivariate relationships captured by our substantive analyses, with loess curves fit by party. In all cases, strong quadratic components appear to be present. As such, we include controls for a squared specification of respective ideal point measures in all substantive regressions reported in the main and supplementary analyses.

\end{document}